\documentclass[aps,pre,floatfix,twocolumn,nofootinbib]{revtex4}
\usepackage[utf8]{inputenc}
\usepackage{dcolumn}
\usepackage[sort&compress]{natbib}
\usepackage{bm}
\usepackage[dvips]{epsfig}
\usepackage{graphicx} \usepackage{amsmath} \usepackage{amssymb}
\usepackage{comment}
\usepackage{mathtools}
\usepackage{adjustbox}
\usepackage{subcaption}
\usepackage{float}
\usepackage{xcolor}
\usepackage{bbm}
\graphicspath{ {Images/} }

\usepackage{systeme}

\newcommand{\E}{\hat{E}}
\renewcommand{\d}{{\rm d}}
\renewcommand{\a}{\alpha}
\newcommand{\BEA}{\begin{eqnarray}}
\newcommand{\EEA}{\end{eqnarray}}

\renewcommand{\comment}[1]{}

\begin{document}

\author{S.G. Babajanyan$^{1,2)}$, E.V. Koonin$^{1)}$ and A.E. Allahverdyan$^{2,3)}$}
\affiliation{$^{1)}$National Center for Biotechnology Information, National Library of Medicine, National Institutes of Health, Bethesda,
MD, USA,  \\
$^{2)}$ Alikahanyan National Laboratory (Yerevan Physics Institute), \\ 2 Alikhanyan Brothers Street, Yerevan 0036, Armenia, \\ 
$^{3)}$ Yerevan State University, 1 A. Manoogian street, Yerevan 0025, Armenia
}

\title{Thermodynamic selection: mechanisms and scenarios}

\begin{abstract}
Thermodynamic selection is an indirect competition between agents feeding on the same energy resource and  obeying the laws of thermodynamics. We examine scenarios of this selection, where the agent is modeled as a heat-engine coupled to two thermal baths and extracting work from the high-temperature bath. The agents can apply different work-extracting, game-theoretical strategies, e.g. the maximum power or the maximum efficiency. They can also have a fixed structure or be adaptive. Depending on whether the resource (i.e. the high-temperature bath) is infinite or finite, the fitness of the agent relates to the work-power or the total extracted work. These two selection scenarios  lead to increasing or decreasing efficiencies of the work-extraction, respectively.  The scenarios are illustrated via plant competition for sunlight, and the competition between different ATP production pathways. We also show that certain general concepts of game-theory and ecology--the prisoner's dilemma and the maximal power principle--emerge from the thermodynamics of competing agents. We emphasize the role of adaptation in developing efficient work-extraction mechanisms.

\end{abstract}

\maketitle

\section{Introduction}
\label{intro}

Thermodynamics studies energy transfer, storage, and usage. It started as a theory of heat engines, drove the Industrial Revolution, and matured at nearly the same time when evolutionary biology emerged. Nowadays, thermodynamics is perhaps the most general phenomenological theory in all of science that applies to all types of systems at all levels of organization. Several attempts have been made to represent various aspects of evolutionary biology, in particular, evolution of populations and ecosystems, within the framework of thermodynamics  \cite{lotka,logofet,odum,sella,dros,vlad,koonin2,eric,mart,ao,jorg}.

Here we develop a thermodynamic approach to selection.  Its main premise is that once organisms (agents) extract work (useful energy) and obey the laws of non-equilibrium thermodynamics in their metabolism \cite{yang}, they can be modeled as heat engines. Agents interact (compete) indirectly, if the extraction goes from the same source. This competition can be represented via game theory, and hence its outcome depends on work-extraction strategies adopted by the agents. Such strategies depend on two parameters: efficiency and power. The second law of thermodynamics states that the efficiency of any heat engine|defined as the ratio of useful extracted energy (work) to the total energy input (heat)|is bound from above by Carnot's efficiency \cite{balian-1,grandy-1,ingo}. But heat engines operating at the maximum efficiency yield zero work per unit of time (zero work power) resulting in the well-known power-efficiency tradeoff \cite{novikov,curzon,broeck,mahler}: the most efficient regime is not powerful, whereas the most powerful regime is not efficient \cite{armen}.

The energy budget of an organism can be described as three main energy currents: input, storage, and output (waste) \cite{gorshkov, jorg, debt1,debt2}. The relationship between these three currents are similar to that in a generalized heat engine: input heat, work (storage), and output heat. Similar to abiotic heat engines, organisms also face the power-efficiency (or speed-fidelity) trade-off. In particular, this trade-off is seen  in molecular machines of cells  \cite{dill1,dill2,angulo,brown}, and also at the level of organism phenotypes \cite{roach,shuster,aledo1,aledo2,hans,spitz,tess}. The power-efficiency trade-off is subject to selection and depends on available energy resources.

Hence, our goal is to explore a physical model for the evolution of the metabolic power-efficiency trade-off, where agents are modeled as heat engines. We do not specify how the extracted work is utilized (reproduction, metabolism, defense, or other functions). Instead, we focus on different strategies (phenotypes) that are available to the agents to extract and store energy. The competition and selection emerge because at least two agents employ the same source (high-temperature bath). There are two general scenarios for such competition, for effectively infinite and for finite|and hence depletable|resources. The quantities relevant for evolution in these two situations are, respectively, the power of work extraction and the stored energy (=total extracted work).

Competition for an infinite resource is analogous to the competition of plants for light. Here the source, i.e. the Sun, acts as a thermal bath providing high temperature photons for the heat engine operation of the photosynthesis. It is not depletable, and yet, there is a competition for a limited energy current reaching the forest surface  \cite{gior,ked,weiner,smith,alpert,mor,ipon,funk,falster,anten,funk2}. Plants can behave differently when facing such competition, from confrontation to avoidance of the competitor \cite{novo,grunt,funk2}. In section \ref{III} we formalize and examine these situations that can have more general relevance in the context of nutrient allocation between cells in multicellular organisms. In particular, we show that the competition leads to increasing the efficiencies consistently with observations. 

Exploitation of a finite source is a dynamical process, since this source is depleted due to the functioning of the agents themselves. We study this process in section \ref{IV}  and show that competition favors heat engines with lower efficiencies. An example of this is the fermentation (aerobic and anaerobic) and respiration pathways of ATP production in yeasts  \cite{shuster,shuster1,maclean,veiga,aledo1,aledo2} and in solid tumor cells \cite{zheng,vander,liberti,hanahan}. Here the ATP production refers to work-extraction and storage \cite{mcclare}. Respiratory ATP production is far more efficient than fermentation, but the speed and hence the power of the fermentation path is greater \cite{voet, melkon,shuster1}. Given the available resources and the presence of competition, cell choose one or the other pathway of ATP production \cite{shuster,shuster1, maclean}. 

Agents competing for a depletable resource alter the common environment similarly to what happens in niche construction theories \cite{lal,old,lala}. Thereby they shape the selection process. Hence, we face a non-trivial game-theoretic situation, where the optimal values of power and efficiency under competition are not unique. However, the environmental changes caused by the behavior of competing agents are ``myopic'', that is, the behavior of the agents is not based on perception of  the global environmental state.

 The common environment of competing agents changes due to the very engine functioning. This fact poses the problem of adaptive (i.e. structure adjusting) {\it versus} non-adaptive agents. This is analogous to the phenotype adaptation that is observed in organisms \cite{ham,ham1,ham2,marek,toloz,roach,mey,stearn,fors,pier}. As seen below, adaptation plays an important role in selection process.

The rest of this paper is organized as follows. Next section defined the heat engine model we employ. Section \ref{III} addresses competition for an infinite resources that amounts to sharing a fixed energy current. Section \ref{IV} studies the competition for a finite resource. We explore this situation via studying two competing agents that can be adaptive or not. Sections \ref{III} and \ref{IV} can be read independently from each other. Both sections employ ideas and techniques from game theory, though no deep preliminary knowledge on this subject is assumed, since we provide the necessary background. We summarize in the last section. All technical derivations are relegated to Appendices. 

\section{Thermodynamic agent (heat engine)}
\label{II}

\subsection{Heat-engine model}

\begin{figure}
    \includegraphics[angle=270, width=7cm]{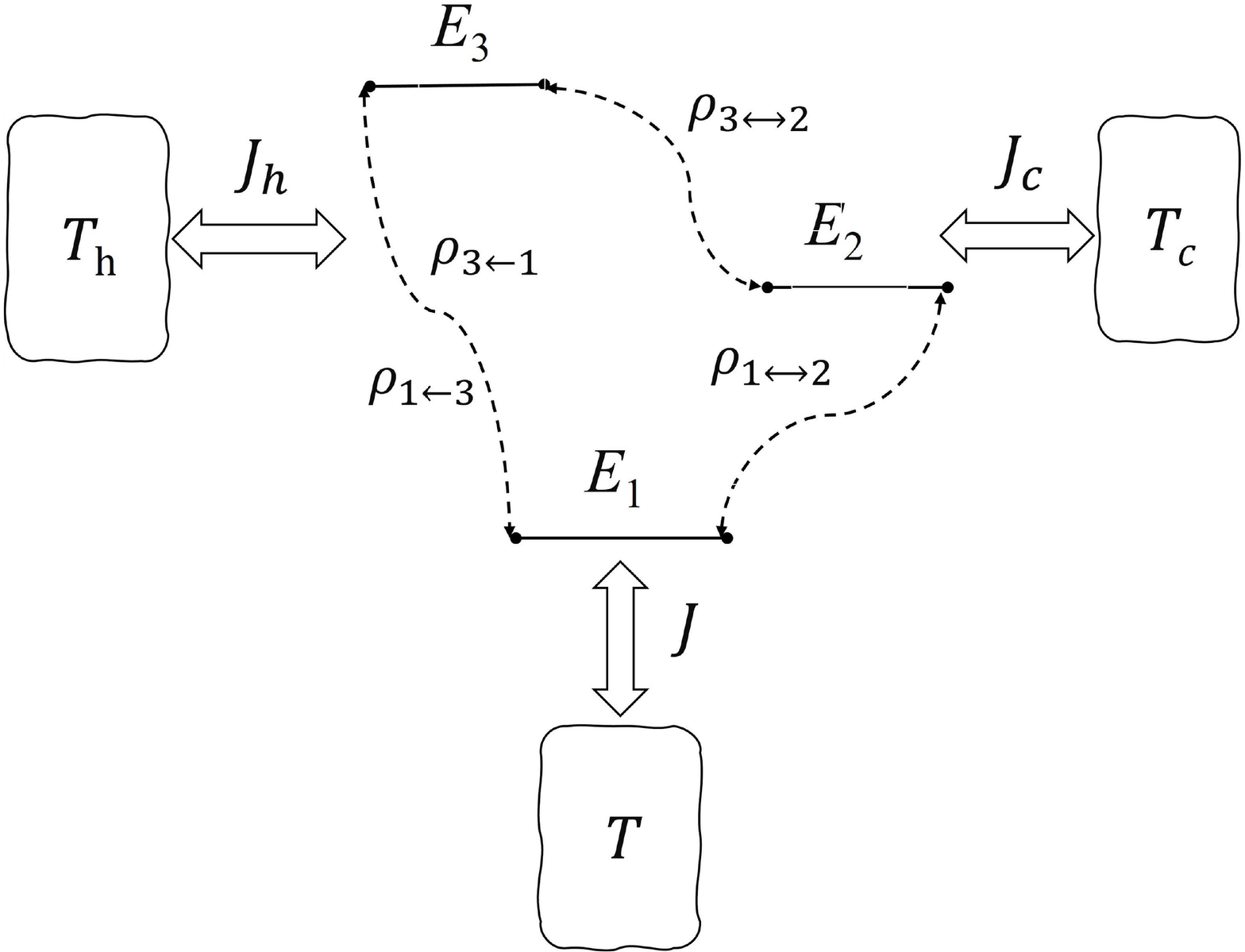}
    \caption{A schematic representation of the heat-engine model (\ref{bo}--\ref{mp}), and related thermal baths that cause transitions between the three engine states with energies $E_{i},\, i=1,2,3$. Now $\rho_{i \leftarrow j}$ is the transition rate from state $j$ to state $i$. 
    $J_{h},\, J_{c}, \, J$ are heat currents from the thermal baths with temperature $T_{h},\,T_{c},\, T$, respectively.} 
\label{engine}
\end{figure}

To model energy extraction and storage, we focused on the minimal thermodynamically  consistent model of a 
heat engine \cite{caplan,broeck}.  For further clarity, we start with the explicit implementation of this model via three-energy level Markov systems attached to different heat baths at different temperatures; see Fig. {\ref{engine}} and Appendix \ref{ap_a} for details. Having an explicit model is essential for clarifying the nature of the involved parameters and the extracted work (stored energy). However, the model will be explored in the high-temperature (linear response) regime, where the implementation details are not essential, and where it is equivalent to linear thermodynamic models employed in biophysics \cite{caplan}. 

The engine has three states $i=1,2,3$. This is the minimal number of states a stationary heat engine can have, because it should be in a non-equilibrium state (i.e. to support one cyclic motion), and has to support three external objects, one work-source and two thermal baths. Each state $i$ has energy $E_i$, such that 
\BEA
\label{bo}
E_{1}=0<E_{2}<E_{3}.
\EEA
Transitions between each pair of different states are caused by the different thermal baths having different temperatures ($T_{\rm h}$, $T_{\rm c}$, $T$) that accordingly provide or accept necessary energies; cf.~Fig.~\ref{engine}. We assume that these thermal baths are in thermal equilibrium states, which means that the transition rates that drive Markov evolution of the engines obey the detailed-balance condition, for example the transition rates between states $\left\{1, 3\right\}$ satisfy the following relation 
\BEA
\label{detal}
\rho_{1 \leftarrow 3}~e^{-\beta_{h}E_{3}}=\rho_{3 \leftarrow 1} ~e^{-\beta_{h}E_{1}},
\EEA
where $\beta_{h}=1/T_{h}$. 
Similar relation holds for the transition rates $\rho_{1 \leftarrow 2}$ and $\rho_{2 \leftarrow 3}$
caused by thermal baths with temperature $T$ and $T_{c}$, respectively.

One temperature is assumed to be infinite \cite{ada}: $\beta=1/T=0$. This bath is then a
work-source. This key point can be explained as follows. First, note that an infinite temperature thermal bath exchanges $\d E$ energy without changing its own entropy, $\d S=\beta\d E=0$, which is a feature of mechanical device (a sources of work) \cite{ada}. Second, if the $T=\infty$-bath spontaneously interacts with any (positive) temperature bath, then the former bath always looses energy. Hence, its energy is freely convertible to any form of heat, as expected from work. Next, we assume $T_{\rm h}>T_{\rm c}$, as necessary for heat engine operation.

In the stationary (but generally non-equilibrium) state the average 
energy of the three-level system is constant
\begin{eqnarray}
\label{n1}
\sum_{i=1}^3\frac{d {p}_i}{d t} E_i=J_{\rm h}+J_{\rm c}+J=0,
\end{eqnarray}
where $p_{i}$ is the probability of finding the system in energy level $E_{i}$. Here $J$ and $J_{n}$ with $n= {\rm h, c}$ are the average energy lost (for ${J},\, J_{n}>0$) or gain (${J},\, J_{n}<0$) by each bath per unit of time. Eq.~(\ref{n1}) is the first law of thermodynamics for a stationary state \cite{callen}.
Now (\ref{n1}) indicates on a perfect coupling between the thermal baths and the three-level system: there is no an energy current standing for irreversible losses within the system; cf.~\cite{voet,stucki,tor,aledo1,angulo,kedem}.

In the stationary state, the energy currents hold
\begin{eqnarray}
\label{n2a}
&& J=\frac{E_2}{{\cal Z}}\rho_{2\leftarrow1}\rho_{1\leftarrow 3}\rho_{3\leftarrow 2}\big(1-e^{(\beta_c-\beta_h)E_3-\beta_c E_2}\big),\\
&& J_{\rm h}= -E_{3} J/E_{2}, ~~ J_{\rm c}=(E_{3}-E_{2})J/E_{2}.
\label{n2}
\end{eqnarray}
where $\cal Z$ is the normalization factor defined by transition rates (see Appendix~{\ref{ap_a}}).

If the system functions as a heat engine, i.e. on average, pumps energy to the work-source, then 
\BEA
J<0, \qquad J_{\rm h}>0, \qquad J_{\rm c}<0,
\label{boa}
\EEA
 Using Eq.(\ref{n2},\ref{n2a}) and the condition  (\ref{boa})  one get the condition for the system to operate as a heat-engine

\begin{eqnarray}
\label{n3}
E_{2}\left[(1-\vartheta)E_{3}-E_{2}\right]>0, ~~ \vartheta\equiv T_{\rm c}/T_{\rm h}=\beta_{\rm h}/\beta_{\rm c}.
\end{eqnarray}
The efficiency of any heat engine is defined as the result (the
extracted work) divided over the resource (the energy coming from the hot bath).
Recalling $E_3>E_2$, the efficiency $\eta$ reads from (\ref{n3}):
\begin{eqnarray}
  \label{n4}
  \eta\equiv\frac{-J}{J_{\rm h}} =\frac{E_{2}}{E_{3}}
  \,\leq\, \eta_{\rm C}\equiv
  1-\vartheta.
\end{eqnarray}
Hence, the efficiency $\eta$ is bounded from the above by the Carnot efficiency
$\eta_{\rm C}$. Eq.~(\ref{n4}) is the second law of thermodynamics for the heat engine efficiency \cite{callen}.

Eqs.~(\ref{n2a}, \ref{n3}, \ref{n4}) demonstrate the power-efficiency trade-off: at the maximum efficiency $\eta=\eta_{\rm C}$ the power $-J$ of the heat engine nullifies. This trade-off is also a general feature of heat engines \cite{mahler}. A clear understanding of this trade-off is one pertinent reason for having an explicit microscopic model of a heat engine.

The work power $-J$ depends on the specific form of the transition rates that enter the detailed balance condition (\ref{detal}) (see Appendix \ref{ap_a}); for example the Arrhenius form of transition rates applies in chemical reaction dynamics \cite{kampen}. Here we work in the high-temperature limit, where the details of rates are not important provided they hold the detailed balance. Now $E_i/T_{\rm c}\ll 1$ and $E_i/T_{\rm h}\ll 1$, but $0\leq \vartheta\leq 1$ in (\ref{n3}) can be arbitrary. In this limit the heat engine power reads via Eq.(\ref{n2a}, \ref{n4}):
\begin{eqnarray}
  \label{eq:1bn}
 - J  = \rho \beta_{\rm c} E_{3}^{2} \eta (1-\vartheta-\eta),
\end{eqnarray}
where $\rho=\frac{1}{{\cal Z}}\, (\rho_{2\leftarrow 1}\rho_{1\leftarrow 3}\rho_{3\leftarrow 2})|_{\beta_{h}=\beta_{c}=0}$ is a constant. 
Eq.~(\ref{eq:1bn}) shows that for a fixed $\vartheta$ the maximum power of $|J|$ of the engine is attained for
\BEA
\eta=\frac{1-\vartheta}{2}=\frac{\eta_{\rm C}}{2}.
\label{mp}
\EEA

\comment{
 Note that for $\beta_{31}=\beta_{32}=0$ we get $\rho_{ij}=\rho_{ji}$ from (\ref{eq:1w}). 
 Hence in this limit we get $p_i=1/3$; cf.~(\ref{eq:3w}). Now $J_{21}\not =0$ in 
 (\ref{eq:1bn}) due to small deviations of $p_i$ from $1/3$ that exist due to a small 
 but finite $\beta_{31}$ and $\beta_{32}=0$.
}

\subsection{Relevance of heat-engines in biology}

Below we shall heuristically apply the heat engine model to ATP production, where the two thermal baths refer to the e.g. glucose (the major reactant of the ATP production), while the work corresponds to the energy stored in ATP, which is metastable at physiological conditions. In this context let us discuss to which extent heat-engine models can be applied to transformations of chemical energy. The standard understanding of the chemical energy stored in certain molecular degrees of freedom is that it is isothermal and is described by the (Gibbs) free energy difference between reactants and products. This coarse-grained description does not tell where (in which degrees of freedom) the energy was stored and how it came out. Detailed mechanisms of such processes are still unclear, e.g. there is a long-standing and on-going debate on how precisely ATP delivers the stored energy during its hydrolysis and how this energy is employed for doing work; see e.g. \cite{mcclare,baker,japan}. However, it is clear that at sufficiently microscopic level all types of stored energy should be related to the fact that certain degrees of freedom are not in the thermal equilibrium with the environment \cite{mcclare}. Indeed, if all degrees of freedom would be thermalized at the same temperature, the second law will not allow any work-extraction \footnote{Even for those cases that seem completely isothermal|e.g. the mixing of different gases at the same temperature and pressure|there is clearly a degree of freedom (difference between gases) that is out of equilibrium.}. It is known that frequently such non-thermalized degrees of freedom can be described by different effective temperatures \cite{jaynes}. Moreover, even a finite non-equilibrium system can (under certain conditions) play the role of a thermal bath, since the dynamics of its subsystem obeys the detailed balance condition \cite{jar}. Thus when describing work-extraction from chemical energy, it is meaningful to assume two different thermal baths, which is in fact the simplest situation of a non-equilibrium system. Modeling work-extraction through different chemical potentials (a situation closer to the standard understanding of the stored chemical energy) is in fact structurally similar to heat-engines \cite{rmp,wang}, also because we work in the high-temperature limit, where many implementation details are irrelevant. In this limit our model is fully consistent with linear equilibrium thermodynamics \cite{caplan,broeck}. Similar models have been widely employed in bioenergetics for modeling coupled chemical reactions, where the passage of heat from higher to lower temperatures corresponds to the down-hill reaction, whereas work extraction corresponds to the up-hill reaction \cite{caplan}.

\section{Competition for fixed current }
\label{III}

Two agents ${1}$ and ${2}$ competing for the same resource can be described as two heat engines attached to the same thermal baths [cf.~(\ref{eq:1bn})]. The sum of the currents coming from the high-temperature bath $T_{\rm h}$ to each heat engine is bound from above [cf.~(\ref{boa})]:
\begin{eqnarray}
\label{fix:1}
J_{\rm 1 \, h}+J_{\rm 2 \,h} \leq A,\qquad A>0,
\end{eqnarray}
where $A$ is a positive constant, and $J_{\rm 1 \, h}$ and $J_{\rm 2 \, h}$ are the currents coming from the same $T_{\rm h}$-bath to each heat engine; cf.~(\ref{boa}).

Using (\ref{eq:1bn}, \ref{n2}) we write constraint (\ref{fix:1}) as 
\BEA
 \label{fix:1a}
&& \eta_1+\epsilon\eta_2\geq a(1-\theta)(1+\epsilon), \qquad \epsilon\equiv E^2_{3 \, 2}/E^{2}_{3 \, 1},\\
&& a\equiv 1-\frac{1}{(1-\vartheta)(1+\epsilon)}\,\frac{A}{E^2_{3 \, 1}}.
\label{aaa}
\EEA
Here $E_{3 \, k}$ is the third (highest) energy level of agent $k=1,2$, $\epsilon$ is the ratio of energy scales of the engines that also determines the difference between ${1}$ and ${2}$. For $\epsilon=1$, the two agents are equivalent, i.e. the situation is symmetric.

We introduce scaled efficiencies $x_k=\eta_k/\eta_{\rm C} \in [0,1]$ for agents $k={1}$ and $k={2}$. Then the work extracted by an agent reads
\BEA
&& -J_{k}=\rho\beta_{c} E^2_{3 \, 1}(1-\vartheta)^2 {\cal W}_k, ~k=1,2,\\
&& {\cal W}_1=x_1(1-x_1),\quad {\cal W}_2=\epsilon x_2(1-x_2).
\label{dora}
\EEA
For $a$ in (\ref{aaa}) we assume 
\BEA
\label{aa}
{1}/{2}<a<1.
\EEA
The first inequality in (\ref{aa}) ensures that the agents cannot simultaneously maximize ${\cal W}_{1}$  and ${\cal W}_{2}$ at $\eta_{1}=\eta_{2}=(1-\vartheta)/2$ (i.e. at $x_{1}=x_{2}=1/2$), which refers to the maximum power regime (\ref{mp}) for both agents.  

The second inequality in (\ref{aa}) means that (\ref{fix:1a}) still allows some $x_{1}<1$  and $x_{2}<1$. Thus, Eq.~(\ref{fix:1}) creates a non-trivial competition between the agents, which translates into the following game theoretical problem
\begin{eqnarray}
\label{fix:2}
&&{\cal W}_1=x_1(1-x_1),\quad {\cal W}_2=\epsilon x_2(1-x_2),
\\
&&x_{1}+\epsilon x_{2} \geq a(1+\epsilon), 
\label{fix:3}
\end{eqnarray}
where $x_{k}$ and ${\cal W}_{k}$ are the actions and payoffs of the agents, respectively.

Thus, the present game has two dimensionless parameters: $a$ and $\epsilon$. Now $a>1/2$ means that the constraint is non-trivial, while $\epsilon$ defines the payoff asymmetry between 1 and 2; cf.~(\ref{dora}).

\comment{
Eq.~(\ref{fix:1}) creates a non-trivial competition between the agents, which translates into the following game theoretical problem (see Appendix \ref{ap_b} for details):
\begin{eqnarray}
\label{fix:2}
&&{\cal W}_1=x_1(1-x_1),\quad {\cal W}_2=\epsilon x_2(1-x_2),
\\
&&x_{1}+\epsilon x_{2} \geq a(1+\epsilon), \quad 1> a > \frac{1}{2},
\label{fix:3}
\end{eqnarray}
where $x_k=\eta_k/\eta_{\rm C} \in [0,1]$ ($k=1,2$) are the scaled efficiencies of (resp.) ${1}$ and ${2}$, and ${\cal W}_{k}$ are the payoffs of the agents found from work-powers $J_{1}$, $J_2$ such that ${\cal W}_{1}/{\cal W}_2=J_{1}/J_{2}$. 
Here $\epsilon=(E_{3 [2]}/E_{3 [1]})^2$ is the ratio of energy scales of the engines that also determines the difference between ${1}$ and ${2}$. For $\epsilon=1$, the two agents are equivalent, i.e. the situation is symmetric.

Eq.~(\ref{fix:2}) implies that ${1}$ and ${2}$ will "strive" to maximize their respective payoffs ${\cal W}_{1}$ and ${\cal W}_{2}$ by taking actions $x_1$ and $x_2$ that hold (\ref{fix:3}) which is inherited from (\ref{fix:1}) via (\ref{n2}). This constraint ensures that the action $x_{k}$ of the agents are coupled and the payoff of each agent depends on the action of the competitor. In the absence of constraint (\ref{fix:3}), maximization of the payoffs ${\cal W}_{k}$ yields $x_{k}=1/2$ ($k=1,2$) that refer to the maximum power regime (\ref{eq:1bn}). Hence, $a > \frac{1}{2}$  in (\ref{fix:3}) guarantees that both agents cannot operate in the maximum power regime (\ref{mp}), whereas $a<1$ ensures that non-trivial competition between the agents remains possible. Thus, a larger $a$ refers to more severe resource constraints.
Now $a=1$, implies $x_{k}=1$ resulting to ${\cal W}_{k}=0$.
}

Game theory offers several concepts of equilibrium that formalize the ill-defined|due to (\ref{fix:3})|notion of ``joint" maximization for the utilities ${\cal W}_1$ and ${\cal W}_2$ \cite{nasho,myerson,luce}. 
For the game considered here, only the equilibrium concepts of Stackelberg and Pareto-Nash are non-trivial, and these are examined below. 

\begin{figure}[!t]
\centering
\includegraphics[width=6cm]{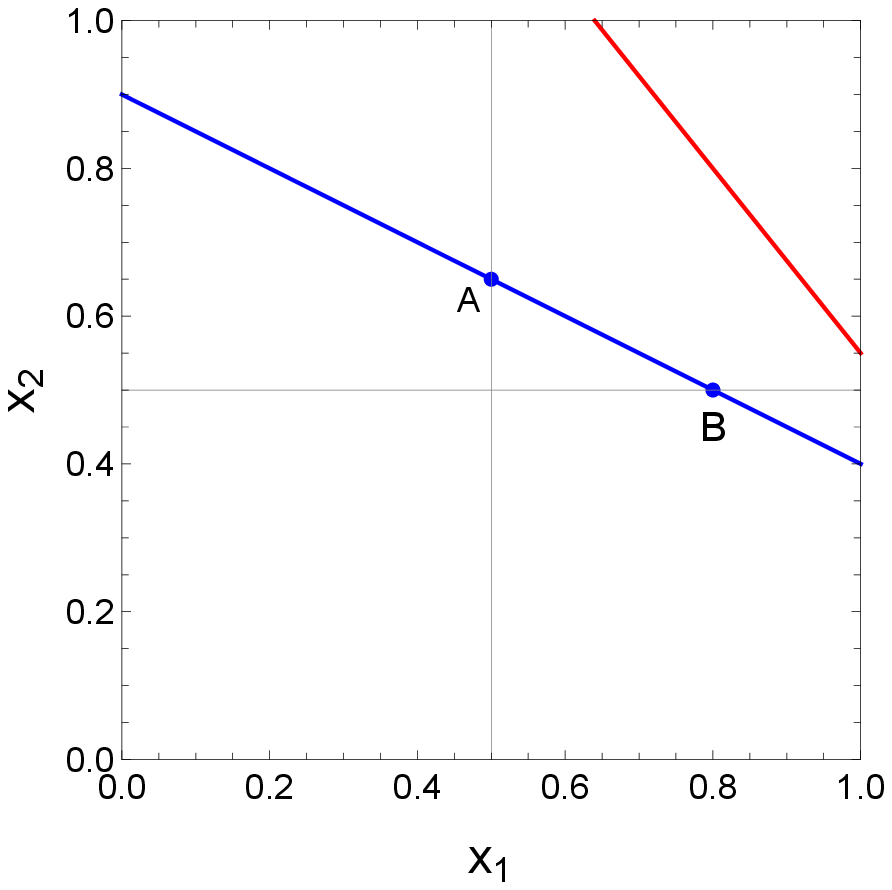}
\caption{
The Pareto line (\ref{nanar}, \ref{nunush}) 
of the game (\ref{fix:2}, \ref{fix:3})
for different values of $a, \, \epsilon$.\\
Blue (lower straight) line is $x_1+\epsilon x_2=a(1+\epsilon)$ with $a=0.6, \, \epsilon=2$. 
The allowed states are located above this line still confined by the square $[0,1]\times[0,1]$. 
The Pareto line is the fraction $[A,B]$ of this line
that lies between intersections with $x_1=1/2$ and $x_2=1/2$; cf.~(\ref{nunush}). The worst outcomes for the players|defined by (\ref{worse}, \ref{dedo1}, \ref{dedo2})|are ${\cal W}^*_{2}\neq 0$ and ${\cal W}^*_{1}=0$. \\
Red (upper straight) line is $x_1+\epsilon x_2=a(1+\epsilon)$ with $a=0.8,\,\epsilon=0.8$. Now the whole red line is the Pareto line. The worse outcomes for the players are ${\cal W}^*_{2}={\cal W}^*_{1}  = 0$.
}
\label{fig:par}
\end{figure}

\begin{figure*}[!t]
\begin{subfigure}{0.45\textwidth}
    \includegraphics[width=7.5cm]{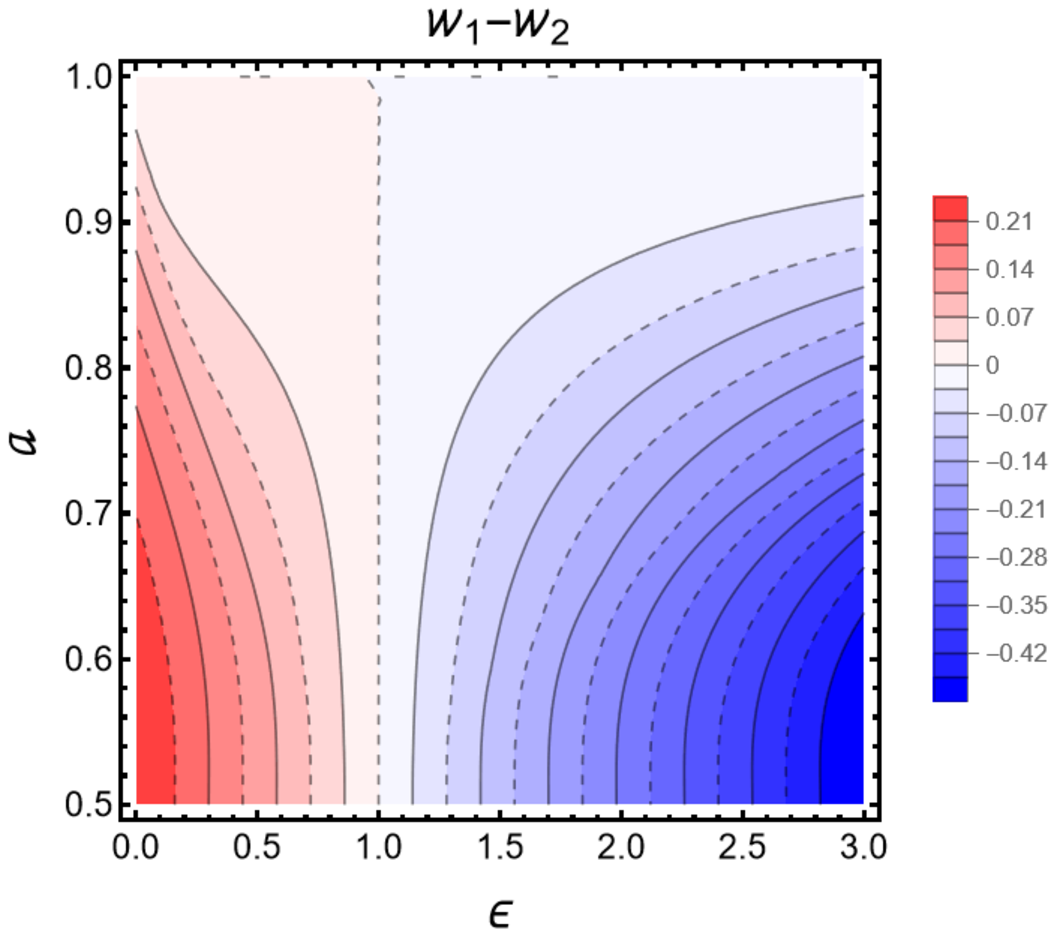}
    \subcaption{}
    \label{fig:01a}
    \end{subfigure}
    \begin{subfigure}{0.45\textwidth}
    \includegraphics[width=7.5cm]{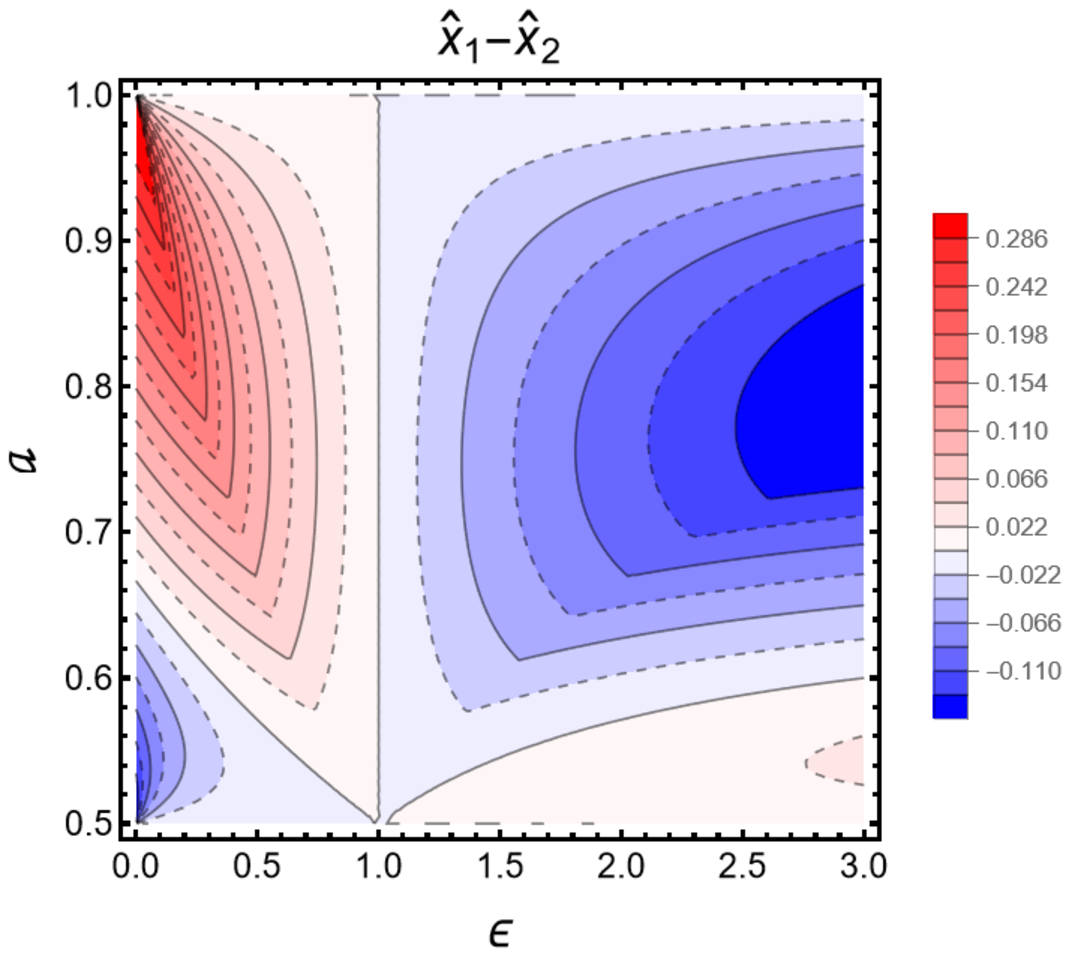}
    \subcaption{}
    \label{fig:01b}
    \end{subfigure}
\caption{The Nash bargaining solution for the game (\ref{fix:2}, \ref{fix:3}). \\ 
(a): ${\cal W}_1-{\cal W}_2$ obtained from (\ref{raaz}) for different $a$ and $\epsilon$. Red (blue) regions refer to ${\cal W}_1-{\cal W}_2>0$ [${\cal W}_2-{\cal W}_1>0$]. Darker red (darker blue) colors refer to larger ${\cal W}_1-{\cal W}_2$ [larger ${\cal W}_2-{\cal W}_1$]. Lines represent contours of fixed values for ${\cal W}_1-{\cal W}_2$. \\
(b): $\hat x_1-\hat x_2$ for different $a$ and $\epsilon$, where $(\hat x_1, \hat x_2)$ are found from
argmax in (\ref{raaz}). Colors follow the above logic.}
\label{fig:01}
\end{figure*}

\subsection{ Stackelberg's equilibrium}

Stackelberg's competition model \cite{myerson,st,we1} is a sequential solution of the above game: the first agent (${1}$) is the leader, since it has the advantage of the first move.  The second agent (${2}$) is the follower that responds to the first move by ${1}$. Hence ${1}$ chooses $x_1=1/2$, which is the unconstrained maximum of ${\cal W}_1$. The best response of ${2}$ to this action reads from the first inequality in (\ref{fix:3})
\BEA
\label{ew}
x_2=\frac{1}{\epsilon}\left[a(1+\epsilon)-\frac{1}{2}\right]\leq 1,
\EEA
where the last inequality amounts to 
\BEA
\label{wew}
a(1+\epsilon)<\frac{1}{2}+\epsilon.
\EEA
There are cases where (\ref{wew}) does not hold even though the second inequality in (\ref{fix:3}) still holds. In such cases, no action by ${2}$ is possible because $ x_2> 1$. Hence the co-existence of two agents is impossible. Satisfying (\ref{wew}) naturally becomes harder for $a \to 1$ and easier for larger $\epsilon$ so that (\ref{wew}) always holds for $\epsilon\to \infty$. Thus, the follower tends to have a sufficiently large $\epsilon$ because otherwise, it is eliminated from the competition. Whenever (\ref{wew}) holds, ${2}$ operates with a higher efficiency than ${1}$, that is, invaders are more efficient \cite{funk,funk2}. For $\epsilon<1$ the work extracted by ${2}$ is smaller than the work extracted by 1: ${\cal W}_2<{\cal W}_1=1/4$; cf.~(\ref{fix:2}). However, ${\cal W}_2>{\cal W}_1$ for $\epsilon>1$ and a sufficiently small $a$, that is, the second agent can extract a larger amount of work at a higher efficiency; cf.~(\ref{fix:2}, \ref{ew}).

\subsection{Pareto line, Nash equilibria and evolutionary stability} 

A pair $(x_1^{[\rm P]}, x_2^{[\rm P]})$ is a Pareto equilibrium of the game if for any allowed $x_1\not=x_1^{[\rm P]}$ and $x_2\not=x_2^{[\rm P]}$ it holds \cite{nasho,myerson,luce}
\BEA
\label{fe1}
{\rm If}\quad {\cal W}_1(x_1, x_2^{[\rm P]})>
{\cal W}_1(x_1^{[\rm P]}, x_2^{[\rm P]}),\\
{\rm then}\quad
{\cal W}_2(x_1, x_2^{[\rm P]})<
{\cal W}_2(x_1^{[\rm P]}, x_2^{[\rm P]}). \\
{\rm If}\quad {\cal W}_2(x_1^{[\rm P]}, x_2)>
{\cal W}_2(x_1^{[\rm P]}, x_2^{[\rm P]}),\\
{\rm then}\quad
{\cal W}_1(x_1^{[\rm P]}, x_2)<
{\cal W}_1(x_1^{[\rm P]}, x_2^{[\rm P]}). 
\label{fe4}
\EEA
Eqs.~(\ref{fe1}--\ref{fe4}) mean that if a change from $(x_1^{[\rm P]}, x_2^{[\rm P]})$
benefits one agent, it implies losses 
for another agent. Now $(x_1^{[\rm P]}, x_2^{[\rm P]})$ is an equilibrium, as far as there is a certain cooperation between the agents preventing any move $(x_1^{[\rm P]}, x_2^{[\rm P]})\to (x_1, x_2^{[\rm P]})$
or $(x_1^{[\rm P]}, x_2^{[\rm P]})\to (x_1^{[\rm P]}, x_2)$ \cite{nasho,myerson,luce}. 
This contrasts the above Stackelberg's solution, where agent ${1}$ chooses the optimal condition for itself, irrespective of what happens to ${2}$.

Eqs.~(\ref{fix:2}, \ref{fix:3}) show how to find the set of Pareto equilibria (i.e. the Pareto line). Note that both ${\cal W}_1=x_1(1-x_1)$ and ${\cal W}_{2}=\epsilon x_2(1-x_2)$ achieve their absolute maxima at (resp.) $x_1=1/2$ and $x_2=1/2$. Hence the Pareto line is the line $x_1+\epsilon x_2= a(1+\epsilon)$ possibly bound by lines $x_1=1/2$ and $x_2=1/2$: 
\comment{\begin{gather}
\label{nanar}
x_1+\epsilon x_2= a(1+\epsilon), \quad 0\leq x_k\leq 1, \quad k=1,2, \\ 
x_2 \in \left[{\rm max}[\frac{1}{2},\,\,\frac{1}{\epsilon}(a(1+\epsilon)-1)], \right.\nonumber \\ \left.{\rm min}[\frac{1}{\epsilon}(a(1+\epsilon)-1),\,\,1]
\right]. 
\label{nunush}
\end{gather}}
\begin{gather}
\label{nanar}
x_1+\epsilon x_2= a(1+\epsilon), \quad 0\leq x_k\leq 1, \quad k=1,2, \\ 
{\rm max}[\frac{1}{2},\,\,
a(1+\frac{1}{\epsilon})-\frac{1}{\epsilon}]\leq x_2\leq {\rm min}[a(1+\frac{1}{\epsilon})-\frac{1}{2\epsilon},\,\,1].
\label{nunush}
\end{gather}
Examples of Pareto lines (\ref{nanar}, \ref{nunush}) are presented in Fig.~{\ref{fig:par}} for different values of model parameters. 

Note that there is a continuum of Pareto equilibria so that additional reasoning is needed to select a unique outcome. Moreover, for this game, all Pareto equilibria are also  Nash equilibria.

Let us discuss this point in more detail, and also outline relations with evolutionary stable strategies. Recall that a Nash equilibrium of the game with payoffs $W_k(x_1, x_2)$ and strategies $x_1$ and $x_2$ for each agent $k={1},\,{2}$ is a pair of strategies $\left(x^{*}_1,x^{*}_2\right)$ such that for any allowed $x_1\not=x_1^*$ and $x_2\not=x_2^*$ \BEA
\label{nash}
W_{1}(x^*_1,x^*_2) \geq W_{1}(x_1, x^{*}_2),~ W_{2}(x^*_1,x^*_2) \geq W_{2}(x^*_1,x_2).
\EEA
Eqs.~(\ref{nash}) mean that for each agent any unilaterial deviation from $\left(x^{*}_1,x^{*}_2\right)$ is not beneficial. If the inequalities in (\ref{nash}) are strict, the Nash equilibrium $(x^*_1,x^*_2)$ is said to be strict. Note that in our problem the payoff of the players depend on the opponent's strategy via constraint (\ref{fix:3}). For ${1}$ the best response strategy ${x}^*_1(x_2)$ to the given strategy $x_2$ of ${2}$ reads: ${x}^*_1(x_2) = 1/2$ if $1/\epsilon~\big(a(1+\epsilon)-1/2)\leq x_2\leq1$; otherwise: ${x}^*_1(x_2)=a~(1+\epsilon)-x_2$. Similar relation holds for ${2}$. The intersection of curves ${x}^*_1$ and ${x}^*_2$ coincides with Pareto line (\ref{nanar}, \ref{nunush}). 

The Nash equilibria in our situation are strict; hence they are also evolutionary stable \cite{eshel}, since no mutant strategy $x_1$ can obtain higher payoff than the resident strategy $x_1^*$: $W_1(x_1,x^*_2)<W_1(x^*_1,x^*_2)$. 

Note that a refinement of evolutionary stability for continuous strategies|known as continuously stable states \cite{eshel}|cannot be applied to our situation. Indeed, the refinement demands from a strict Nash equilibrium to be also a local maximum of the payoffs. However, the notion of a local maximum cannot apply to our situation, since the payoff functions are discontinuous on the Pareto line \cite{cress}, i.e. they are not defined bellow the Pareto line in Fig.~\ref{fig:par}. Further 
consideration of evolutionary stability demands implementing mixed strategies which goes out of our present scope.

\subsection{Worst outcomes}

For any game it is relevant to know what worst thing agent 2 can do for 1, at any expense for 2. Besides describing hostility, worst outcomes will be relevant for bargaining solutions below. The worst outcome ${\cal W}^*_k$ for agent $k=1,2$ read: 
\BEA
\label{worse}
{\cal W}^*_1={\rm max}_{x_1}{\rm min}_{x_2} [\, {\cal W}_1\, ], ~
{\cal W}^*_2={\rm max}_{x_2}{\rm min}_{x_1} [\, {\cal W}_2\, ].
\EEA
 To find out ${\cal W}^*_1$, note from (\ref{fix:3}) that any choice $x_1$ of ${1}$ should hold $x_1\geq a(1+\epsilon)-\epsilon x_2$. If ${2}$ arranged its action $x_2$ such that $a(1+\epsilon)-\epsilon x_2\geq 1$, then there is no choice to be made by ${1}$. For such cases we shall define its utility ${\cal W}_1=0$, and this is clearly the worst outcome for ${1}$. Now $a(1+\epsilon)-\epsilon x_2\geq 1$ or equivalently $[a(1+\epsilon)-1]/\epsilon\geq x_2$ will work provided that $a(1+\epsilon)-1\geq 0$. Hence, if the latter condition holds we get ${\cal W}^*_1=0$. Otherwise, if $a(1+\epsilon)-1< 0$, then ${2}$ achieves ${\rm min}_{x_2} [\, {\cal W}_1\, ]$ when the bound $x_1\geq a(1+\epsilon)-\epsilon x_2$ is possibly tight, i.e. for $x_2=0$, and then $x_1= a(1+\epsilon)={\rm argmax}_{x_1}{\rm min}_{x_2} [\, {\cal W}_1\, ]$. 
 Calculating ${\cal W}^*_2$ in the same way we get 
\BEA
\label{dedo1}
{\cal W}^*_1=a(1+\epsilon)\,{\rm max}[0,1-a(1+\epsilon)], \\
{\cal W}^*_2=a(1+\epsilon^{-1})\,{\rm max}[0,\epsilon-a(1+\epsilon)].
\label{dedo2}
\EEA
Note that ${\cal W}^*_{1}{\cal W}^*_{2}=0$: (\ref{dedo1}, \ref{dedo2}) imply ${\cal W}^*_{1}{\cal W}^*_{2}>0$ if $\frac{a}{1-a}<\epsilon<\frac{1-a}{a}$, which is impossible since $1\geq a\geq \frac{1}{2}$.

The fact of e.g. ${\cal W}^*_{1}>0$ is important, since it means that ${1}$ cannot be eliminated from the competition. Note that
(\ref{dedo1}, \ref{dedo2}) are not simply the separate global minima of $({\cal W}_{1},{\cal W}_{2})$, i.e. generally $({\cal W}^*_{1},{\cal W}^*_{2})\not=(0,0)$. Still we get $({\cal W}^*_{1},{\cal W}^*_{2})=(0,0)$ for $\epsilon=1$.

\subsection{Fair allocation and the Nash bargaining solution}

Bargaining solutions assume additional cooperativity between the agents|that relates to fair division of resources|thereby looking for a unique choice within the Pareto line (\ref{nanar}, \ref{nunush}) \cite{nasho,roth,myerson,luce}. We shall work with the Nash bargaining solution \cite{nasho,roth,myerson,luce}.  An advantage of this over other bargaining solutions is the consistency of underlying axioms with thermodynamic processes \cite{we2}.

The Nash bargaining solution $(\hat x_1,\hat x_2)$ is found from (\ref{dedo1}, \ref{dedo2}) by maximizing the geometric mean of ${\cal W}_{1}-{\cal W}^*_{1}$ and ${\cal W}_{2}-{\cal W}^*_{2}$ \cite{we2,roth,nasho}:
\BEA
\label{raaz}
{\rm max}_{x_1,x_2}\left[ \, ({\cal W}_{1}(x_1)-{\cal W}^*_{1})({\cal W}_{2}(x_2)-{\cal W}^*_{2})\,\right].
\EEA
The meaning of (\ref{raaz}) is that both 1 and 2 (in a sense) simultaneously maximize their increments over the worst outcomes. Note that only one of ${\cal W}^*_{1}$ and ${\cal W}^*_{2}$ can be non-zero, that is, one of the agents will be competed out in the worst cases; see Appendix \ref{ap_b}.

Eq.~(\ref{raaz}) implies that for $\epsilon=1$ (the same energy scales) the maximizers $\hat x_1$ and $\hat x_2$ of (\ref{raaz}) hold $\hat x_1=\hat x_2=a$. Thus, the resources are allocated evenly between equivalent agents. More generally, (\ref{raaz}) predicts $0.5< \hat x_k<1$ and $k=1,2$, that is, none of the agents is eliminated, and none of them operates at the maximum power. Another implication of (\ref{raaz}) is that 
\BEA
{\rm sign}[{\cal W}_{1}-{\cal W}_{2}]={\rm sign}[1-\epsilon],
\label{boto}
\EEA
i.e. for $\epsilon<1$, ${2}$ extracts less work than ${1}$; see Fig.~\ref{fig:01}, where
${\cal W}_{1}-{\cal W}_{2}$ is  presented for various $\epsilon$ and $a$. It is seen that the difference in the obtained resources of the agents increases with the available resource level (i.e. with decreasing $a$, cf. (\ref{fix:3})). As shown by numerical solution of (\ref{raaz}), no simple relation exists for $\hat x_1-\hat x_2$; see Fig.\ref{fig:01}. Indeed, acquiring more resource is possible either with higher efficiency level (left-top red region) or low efficiency level (left-bottom blue region).  

\subsection{Outlook} 

Plants compete for sunlight. They evolved various strategies for that: horizontal growth, vertical growth, shade tolerance {\it etc}) \cite{gior,ked,weiner,smith,alpert,mor,ipon,funk,falster,anten,funk2}. Photosynthes is a heat engine operating between two thermal baths, the hot photon bath generated by the Sun and the cold bath of the ambient environment. The sunlight is not an exhaustible source (in contrast to food), but plants growing on the same territory can appropriate different portions of the sunlight current depending on their competing abilities and strategies. 

We can thus apply the heat-engine model to a single plant. The competition between two plants (i.e. two heat-engines) will be described via (\ref{fix:1}--\ref{dora}). Now (\ref{fix:1}, \ref{fix:3}) correspond to constraints that 
determine the competition between two plants (agents): a larger $a$ signifies stronger competition, while the minimal value $a\to 1/2$ of $a$ means no competition. Once $\epsilon$ determines the agents asymmetry between 
the work-currents [cf.~(\ref{dora})], it is natural to relate $\epsilon$ to the ratio of the full active leaf area of each plant: $\epsilon>1$ means that for agent ${2}$ this area is larger. 

Plants can have two basic competition strategies: 

-- ({\it 1}) direct competition, where ${1}$ (leader) is the native plant, and ${2}$ (follower) invades the already established canopy. 

--({\it 2}) Avoidance of competition, where some compromise on resource sharing is reached. This regime can refer to the fair solution discussed in (\ref{nanar}--\ref{boto}). 

Now ({\it 1}) corresponds to Stackelberg's solution (\ref{ew}). As follows from (\ref{wew}), the invader survives if its active leaf area is larger than that of the native plant ($\epsilon>1$), and it operates with a higher efficiency. Both these features are compatible with the observations on plant competition. 
Examples of invaders that hold these features  are {\it Ligustrum Sinense} \cite{mor}, and {\it Schinus molle} \cite{ipon}. Also, there is a general agreement that a high resource environment (lower $a$) is more vulnerable to invasion than a lower resource one \cite{gior,alpert,funk,funk2}, which is compatible with (\ref{ew}, \ref{wew}). Note that more general models should, in particular, account for the fact that the payoff 
${\cal W}_{k}$ of the agents in the competition non-linearly depend on the ratio of energy scales $\epsilon$.
Indeed, generally the light obtained by a plant does not depend linearly on its size \cite{weiner,ked,gior,mor}.

Another application of this formalism relates to resource sharing in a multi-cellular organisms,
where different cells rely on constant nutrients supply. This assumes cooperation between cells and fair allocation of nutrients based on the resource demand of the cells; hence ({\it 2}) applies. The expensive germ line hypothesis is an example of resource allocation problem between germ and somatic cells in multicellular organisms \cite{kirk,mak,kirk2,chen}. Germ cells require more nutrients, resulting in a clearly asymmetrical situation as in (\ref{fix:2}, \ref{fix:3}). On the other hand, the lack of resources for somatic cells could cause senescence \cite{kirk2}. Hence, the resource allocation between germ and somatic cells can be considered as a bargaining process over the available resources. Under a shortage of resources, the organism would distribute the resources between germ and somatic cells more evenly, as can be seen in Fig.\ref{fig:01}, where $|{\cal W}_1-{\cal W}_2|$ becomes smaller for larger $a$. 

\comment{
For example, according to the ``expensive germ line'' hypothesis, maintenance and repair of germ cells are more costly than for somatic cells \cite{kirk,kirk2,chen,mak}. Hence dividing the resource between germ and somatic cells is a non-trivial problem; e.g. the lack of resources for somatic cells could cause senescence.}

\section{Competition for depletable resources}
\label{IV}

\subsection{Resource depletion}

Above we assumed that the thermal baths are very large, hence their temperatures do not change during the heat engine functioning. Now we focus on finite baths that are still large enough so that their temperature(s) changes on time scales that are much longer than the relaxation time of the engine. 
Because the hotter bath is the source, we shall assume that its temperature $T_{\rm h}$ decreases as the result of the work-extraction. For simplicity, the other two baths will be held at a constant temperature. 

Eq.~(\ref{eq:1bn}) shows that once $T_{\rm h}$ decreases and gets closer to $T_{\rm c}$ the heat engine will stop functioning at $1-\vartheta=E_2/E_3$. This is still compatible with $\vartheta<1$, i.e. the baths are still not in equilibrium, but this non-equilibrium cannot be employed by the engine for work-extraction. 

The slow dynamics of $T_{\rm h}$ can be deduced from the formula of equilibrium heat-capacity,
which governs the change of temperature for a body at equilibrium given the change of its internal energy \cite{callen} [cf.~\cite{bjarne,berry}]:
\begin{eqnarray}
\label{eq:1c}
\frac{d T_{\rm h}}{d t}\equiv \dot{T}_{\rm h}=-C_{\rm h}^{-1}J_{h}
\end{eqnarray}
where $C_{\rm h}$ is the constant heat capacity of the bath. Indeed, $C_{\rm h}$ is large, since it scales with the number of the bath degrees of freedom. On the other hand, $J_{\rm h}\sim 1$, hence $\frac{d T_{\rm h}}{d t}$ is small, i.e. $T_{\rm h}$ changes slowly. 

\subsection{The setup }

For simplicity, we assume that $E_3$ and $\rho$ are the same for both agents: $E_{3 \, 1}=E_{3 \, 2}=E_{3}$ and $\rho_{1}=\rho_{2}$; cf.~(\ref{eq:1bn}). Under these assumptions, the dimensionless power $\hat J_{k}$ of each agent ($k=1,2$) and the  time evolution of temperature ratio $\vartheta$ are given by [cf.~(\ref{eq:1bn}, \ref{n4})]
\begin{eqnarray}
\label{urd33}
&&\hat J_{k} \equiv \frac{J_{k}}{\mu C_{\rm h}} =-\eta_k(1-\vartheta-\eta_k),\quad \eta_k=\frac{E_{2 \, k}}{E_3},   \\
&&\dot\vartheta=2\mu\vartheta^2(1-\vartheta-\bar \eta),\qquad \bar \eta\equiv(\eta_1+\eta_2)/2,
\label{urd44}
\end{eqnarray}
where (\ref{urd44}) is found from (\ref{eq:1c}), $\eta_{k}$ are efficiencies of agents, and $\mu\equiv C_{\rm h}^{-1}\beta_{\rm c}^2\rho E_{3}^2$ is a constant.
We will compare different types of heat engines that employ the same depletable source and compete with each other aiming to increase the stored energy (= extracted work). The heat engine power (\ref{urd33}) and its time-integral (stored energy) describe (resp.) short- and long-term advantages of the agent. 

We distinguish adaptive and non-adaptive agents in the competition for the same depletable source. The internal structure (difference in energy levels) of non-adaptive agents remains fixed during the time, i.e $\eta_{k}= {\rm const},~ k=1,2$ are time-independent. In contrast, adaptive agents continuously tune their time dependent efficiency to the source depletion:
\BEA
\eta_{k} =\alpha_{k}\eta_{\rm C}=\alpha_k(1-\vartheta(t)). 
\label{ada}
\EEA
The rationale of choosing the specific adaptation protocol (\ref{ada}) comes from the locally equilibrium thermodynamics; e.g. $\alpha_k=1/2$ describes the local maximum power according to (\ref{mp}).
Appendix \ref{ap_c} discusses in detail the energy stored by a single agent (adaptive or not).

Adaptation to environmental changes (phenotypic flexibility) is an intrinsic capacity of all organisms \cite{ham,ham1,ham2,marek,toloz,roach,mey,stearn,fors,pier}. Adaptive processes are also involved in niche-construction \cite{lal,old,lala}, where competing agents shape the selection process by altering their own and competitors' environment. In view of (\ref{ada}) we focus on myopic adaptation, where agents adapt to environmental changes caused by their activity, but do not shape the environment beforehand (that is, do not construct new niches). 

We emphasize that (\ref{ada}) is mostly a phenomenological description of adaptation processes. We do not explore internal mechanisms of adaptation, and in particular, we do not account for the full energy cost of maintaining the adaptation; see \cite{ada} in this context. The main advantage of (\ref{ada}) (to be explained below) is that allows the heat engine to work till the final depletion of the source, i.e. till $\theta\to 1$; cf.~the paragraph before (\ref{eq:1c}). 

\subsection{Competition between non-adaptive agents}
\label{arno}

\begin{figure}[!ht]
\includegraphics[width=7cm]{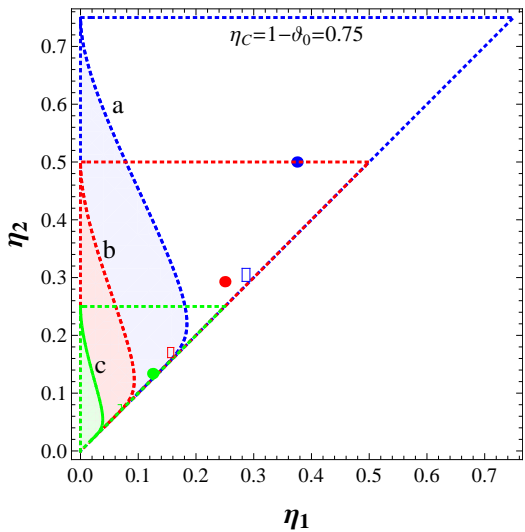}
\caption{Two non-adaptive agents ${1}$ and ${2}$ compete for a depletable source. Their stored energies (payoffs) are given (resp.) by (\ref{na1}) and (\ref{na2}), while the work-extracting strategies are parametrized by efficiencies $\eta_{1}$ and $\eta_{2}$. According to (\ref{n4}, \ref{brun}), $\eta_{1}$ and $\eta_{2}$ hold $1-\theta_0>\eta_{2}>\eta_{1}$. Three situations are shown on the same graph. For $\vartheta_{0}=0.25$ blue-dotted lines delimit the domain $0.75>\eta_{2}>\eta_{1}>0$, while the blue-shadowed domain indicates on $W_{2}(\eta_{1},\eta_{2})>W_{1}(\eta_{1},\eta_{2})$. The rest of the domain delimited by blue-dotted curves refers to $W_{2}(\eta_{1},\eta_{2})<W_{1}(\eta_{1},\eta_{2})$. Likewise, green and red lines and shadowed domains refer to $\vartheta_{0}=0.5$ and $\vartheta_{0}=0.25$, respectively. 
Thus on curves a, b and c we have $W_{2}(\eta_{1},\eta_{2})=W_{1}(\eta_{1},\eta_{2})$ for (resp.)
$\vartheta_{0}=0.25, 0.5, 0.75$. The blue, red, green points refer to $\eta_{1}=\frac{1-\vartheta_{0}}{2}$ and $\eta_{2}=1-\sqrt{\vartheta_{0}}$ for (resp.) $\vartheta_{0}=0.25, 0.5, 0.75$.\\
The blue, red, green rectangles illustrate the outcomes of emergent prisoner's dilemma subgame (\ref{cycle}-- \ref{prison2}) obtained through the best response analysis of the stored energies (\ref{na1}, \ref{na2}). }   
\label{fig:1}
\end{figure}

Let us now consider competition of two non-adaptive agents (${1}$ and ${2}$) attached to the same baths. We can take without loss of generality
\BEA
\label{brun}
\eta_1\leq \eta_2,
\EEA
Hence ${2}$ will have a finite working time determined from (\ref{urd44}) via $\eta_2=1-\vartheta(\tau_2)$. We assume that for $t>\tau_2$, ${2}$ is detached from the baths. Otherwise, this agent will not function as a heat engine. Hence the competition for the same source brings in a finite consumption time of functioning as heat engine. 

For a given initial value $\vartheta(0)=\vartheta_{0}$, define the stored energies for (resp.) ${1}$ and ${2}$:
\begin{eqnarray}
\label{na1}
&& W_{1}(\eta_{1},\eta_{2}) = \int_0^{\tau_2} {\rm d}t \, |\hat J_{1}|+ \int_{\tau_2}^{\infty} {\rm d}t \, |\hat J^{'}_{1}|,\\
&& W_{2}(\eta_{1},\eta_{2}) = \int_0^{\tau_2} {\rm d}t \, |\hat J_{2}|.
\label{na2}
\end{eqnarray}
$W_{1}(\eta_{1},\eta_{2})$ in (\ref{na1}) is composed of two terms. The first term in (\ref{na1}) is the stored energy for time interval $\tau_{2}$, i.e. during the competition. The second term is evaluated once ${2}$ stops work-extraction and ${1}$ operates alone; see Appendix \ref{ap_d}. Note that the functions $|\hat J_{1}(t)|$ and $|\hat J^{'}_{1}(t)|$ are different due to (\ref{urd44}) and the finite working time $\tau_{2}$. $|\hat J^{'}_{1}(t)|$ is given by (\ref{urd33}), and found from (\ref{urd44}) where evaluation starts at $\vartheta(\tau_{2})=1-\eta_{2}$.  

For a single agent in the absence of competition, the optimal efficiency of the heat engine is determined from maximizing the stored energy (time-integrated power) [see Appendix \ref{ap_c}]
\BEA
\eta=1-\sqrt{\vartheta_0}. 
\label{ca}
\EEA
Eq.~(\ref{ca}) is formally similar to the Curzon-Ahlborn efficiency of heat engines \cite{novikov,curzon,broeck}, but the meaning of (\ref{ca}) is different, because (\ref{ca}) follows from the consideration of the full extracted work from a finite bath. We emphasize that (for a single non-adaptive agent) the optimal efficiency (\ref{ca}) is strictly smaller than the maximal (Carnot) efficiency $1-{\vartheta_0}$; cf.~(\ref{n4}). 

Fig.\ref{fig:1} shows the competition results described by (\ref{na1}, \ref{na2}) for two non-adaptive agents ${1}$ and ${2}$. The energy-storing strategies of each agent are parametrized by their efficiencies $\eta_1$ and $\eta_2$; recall (\ref{brun}). It is seen that when competing with the efficiency $\eta_{1}=\frac{1-\vartheta_{0}}{2}$ at the maximal initial power [cf.~(\ref{mp})], it is never meaningful for ${2}$ to employ a larger efficiency, i.e. $\eta_{1}=\frac{1-\vartheta_{0}}{2}$ wins over all $\eta_2>\eta_{1}=\frac{1-\vartheta_{0}}{2}$; see Fig.\ref{fig:1}. In particular, $\eta_2=1-\sqrt{\vartheta_0}$ is not anymore optimal in contrast to the single-agent case (\ref{ca}).
However, when competing against $\eta_{2}=\frac{1-\vartheta_{0}}{2}$ it is beneficial for ${1}$ to employ certain smaller efficiencies $\eta_1<\eta_{2}=\frac{1-\vartheta_{0}}{2}$. 

\subsection{Best responses and emergent prisoner's dilemma}

\subsubsection{Best response cycle}

Once $W_1$ and $W_2$ in (\ref{na1}, \ref{na2}) depend on both $\eta_{1}$ and $\eta_{2}$, the notion of the optimal efficiency is to be studied via game theory. Here the stored energies $W_1\left(\eta_{1},\eta_{2}\right)$ and $W_2\left(\eta_{1},\eta_{2}\right)$ are payoff functions of (resp.) agents ${1}$ and ${2}$, and the choice of $\eta_{1}$ and $\eta_{2}$ (strategy profile) refers to their actions. 

We can determine the best responses of this game that eventually leads to Nash equilibrium. Starting from an arbitrarily initial point $\eta_{1}^{[0]}$ an iterative process $(\eta_{1}^{[0]}, \eta_{2}^{[0]}, \eta_{1}^{[1]}, \eta_{2}^{[1]}, ....)$ of successive best responses is defined recalling (\ref{na1}, \ref{na2}) and (\ref{brun}): 
\BEA
\label{mo1}
&& \eta_{2}^{[k]}={\rm argmax}_{\eta\geq \eta_{1}^{[k]} }[W_{2}(\eta_{1}^{[k]},\eta)], ~~k\geq 0,~~\\
&& \eta_{1}^{[k+1]}={\rm argmax}_{\eta\leq \eta_{2}^{[k]}} [W_{1}(\eta,\eta_{2}^{[k]})], 
~~k\geq 0.~~
\label{mo2}
\EEA
This iteration converges to a 4-cycle that is independent from the initial point $\eta_{1}^{[0]}$:
\BEA
\label{cycle}
&& \eta'_1\to \eta'_2\to \eta''_1\to \eta''_2\to \eta'_1,\\
&& \eta'_1< \eta'_2= \eta''_1< \eta''_2,
\label{lola}
\EEA
where $\eta'_1\to \eta'_2$ means that $\eta'_2$ is the best response to $\eta'_1$. 
Hence the agents can apply the following pairs of strategies: 
\BEA
\label{u1}
(\eta'_{1},\eta'_{2}), \quad (\eta''_{1},\eta'_{2}), \quad
(\eta''_{1},\eta''_{2}), \qquad (\eta'_{1},\eta''_{2}). 
\EEA
We emphasize that the efficiency values in (\ref{lola}) depend only on $\vartheta_0$ and on the initial assumption $\eta_2\geq\eta_1$ in (\ref{na1}, \ref{na2}). For example, at $\vartheta_0=0.5$ we have $\eta'_1=0.151$, $\eta'_2= \eta''_1=0.161$, and $\eta''_2=0.179$. Hence, the long-time best response in this game is not a unique Nash equilibrium, but rather the 4-cycle (\ref{lola}). For $\vartheta_0\to 0$ (potentially large resources), we get that all efficiencies in (\ref{lola}) converge to $2/3$. In contrast, for $\vartheta_0\to 1$ they all converge to zero. Fig.~\ref{fig:1} shows the location of (\ref{u1}) on the $(\eta_1\leq\eta_2)$ plane for several values of $\vartheta_0$; see rectangles in Fig.~\ref{fig:1}. It is seen that they are close to the $\eta_1=\eta_2$ line. 

Hence, the best response dynamics converges to the limit cycle. This is an asymptomatically stable state, because any deviation converges back to the cycle. Note that the above sequential best response algorithm does not ensure that all Nash equilibria of the game are found. Indeed, cycles of best response are widely observed in generic two-player dynamic games \cite{pangal}. For adaptive agents studied below, there is no limit cycle due to the existence of global maxima of the payoff functions.

\subsubsection{Emergent prisoner's dilemma}

Now note that
\begin{align}
\label{prison}
 W_i(\eta_{1},\eta_{2}) < W_{i}\left(1-\sqrt{\vartheta_0},1-\sqrt{\vartheta_0}\right),~i=1,2,
\end{align}
where $(\eta_{1},\eta_{2})$ assumes any of 4 (i.e. any of the allowed) pair in (\ref{u1}). 
Eq.~(\ref{prison}) is non-trivial, because it shows that the joint application of the best response strategies (\ref{lola}) loses to the joint application of the efficiency $1-\sqrt{\vartheta_0}$ that is optimal for a single agent; cf.~(\ref{ca}). 

However, this feature of $1-\sqrt{\vartheta_0}$ is unstable due to
\begin{align}
\label{prison1}
& W_1\left(\eta, 1-\sqrt{\vartheta_{0}}\right)>W_{1}\left(1-\sqrt{\vartheta_0},1-\sqrt{\vartheta_0}\right),\\
\label{prison2}
& \eta=\eta'_1, \eta''_1.
\end{align}
Eq.~(\ref{prison1}) means that switching from $1-\sqrt{\vartheta_{0}}$ to $\eta'_1$ or $\eta''_1$ is beneficial for ${1}$. Now ${2}$ responds to this switch in the best way, and the agents find themselves within actions (\ref{u1}) that are worst than $\left(1-\sqrt{\vartheta_0}, 1-\sqrt{\vartheta_0}\right)$ due to
(\ref{prison}). 

This situation does resemble the prisoner's dilemma \cite{shuster, we1, myerson,aledo2,hof}, where the agents following by best response strategies end up in the worse situation compared with the cooperative behavior, which for our case refers to (\ref{prison}). We emphasize that, in contrast to the standard prisoner's dilemma, here the best-response strategies are not unique and amount to 4-cycle (\ref{cycle}); see rectangles in Fig.{\ref{fig:1}}. Note that the 4-cycle also contains a step with equal efficiencies $\eta'_2= \eta''_1$. 

\subsection{Adaptive agents}

\subsubsection{Competition between two adaptive agents}

Let us consider now the competition between adaptive heat engines (\ref{ada}). 
In the absence of competition (i.e. for a single agent), the optimal value of $\alpha$ in (\ref{ada}) obtained from maximizing the total stored energy (extracted work) is close to the maximal Carnot efficiency (\ref{n4}), i.e. $\alpha \to 1$ in (\ref{ada}); see Appendix \ref{ap_c}. Indeed, the single adaptive agent (\ref{ada}) works as long as $\vartheta \neq 1$ (till the full depletion of the resource). Hence $\alpha \to 1$ minimizes the losses in the energy storage process. Note the difference with (\ref{ca}), where the optimal efficiency for the non-adaptive agent was strictly smaller than the Carnot efficiency. In this sense the usual premise of thermodynamics that the maximal efficiencies would be useful applies to the (single) adaptive agent, which
implies advanced functioning mechanisms, as compared to the non-adaptive situation.

For the competition of two adaptive agents ${1}$ and ${2}$ with $\alpha_1$ and $\alpha_2$ in (\ref{ada}), we get for stored energies [see Appendix \ref{ap_d}]
\begin{eqnarray}
\label{x6}
&&W_k(\alpha_{1},\alpha_{2})=\int_0^\infty {\rm d}t\, |\hat J_{k}(t)| \nonumber\\
&&=\frac{\alpha_{k}(1-\alpha_{k})}{2 \mu (1-\bar\alpha)}\bigg(\frac{1}{\vartheta_0}-1+\ln{\vartheta_{0}}\bigg),
\quad k=1,2,
\end{eqnarray}
where $\vartheta_0$ is the initial ratio of temperatures, and $\bar \alpha\equiv (\alpha_{1}+\alpha_{2})/2$.
Eq.~(\ref{x6}) shows that the agent with $\alpha_k$ closer to $1/2$ wins. Now (\ref{ada}) and (\ref{mp}) imply that $\alpha_k=1/2$ is (locally) the maximum power regime. Note that similar maximum-power regimes were proposed as an operating principle for ecological and biological systems (including living organisms) \cite{odum}. Here, we see that the optimality of  maximum-power regime is closely related to the adaptation abilities of agents. 

Adaptive agents do face the prisoner's dilemma, but the situation here is simpler than that discussed in (\ref{cycle}-- \ref{prison2}). If $\a_1=\alpha_{2} \to 1$, then $\frac{1-\alpha_{i} }{1-\bar\alpha}\to 1$ in (\ref{x6}), and each one will store more energy compared to the case when both operate at the maximum power $\alpha_{k}=1/2$. This situation, again, is unstable: if, for example, the first agent switches to the maximum power regime  $\alpha_{1}=1/2$ (with the second agent working at $\alpha_2\to 1$), then the first agent will store more energy.

\subsubsection{Adaptive agent competing with a non-adaptive one}

\begin{figure}[t]
\centering
\includegraphics[width=6cm]{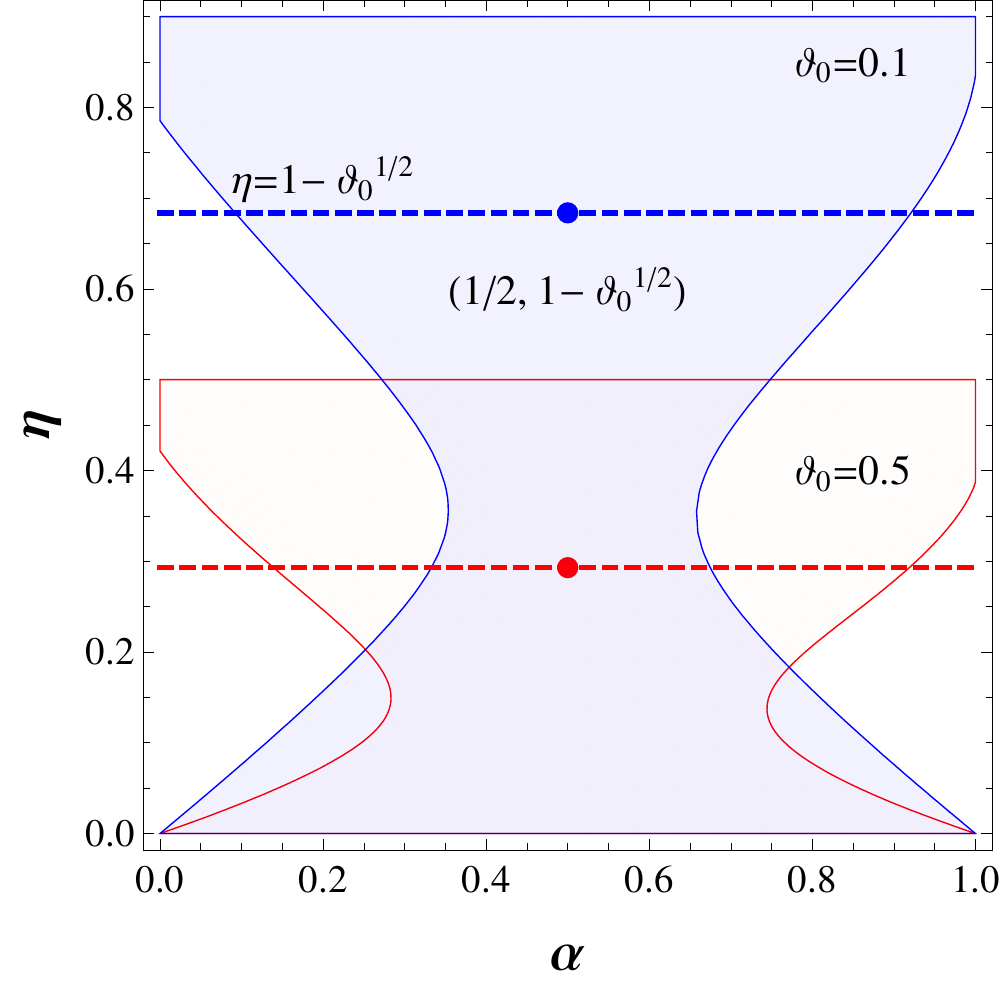}
\caption{Competition between adaptive agent with efficiency $\eta_{ad}=\a(1-\vartheta(t))$ for $\a \in [0,1]$ and a non-adaptive agent with fixed efficiency $\eta\in[0,1-\vartheta_{0}]$; cf.~(\ref{ada}, \ref{n4}). The initial temperature ratios are $\vartheta_{0}=0.1$ (blue) and $\vartheta_{0}=0.25$ (red).  
In shaded regions the adaptive agent extracts more total work than the fixed-$\eta$ agent. Dotted curves show the line $\eta=1-\sqrt{\theta_0}$ for $\vartheta_{0}=0.1$ (blue) and $\vartheta_{0}=0.25$ (red).  
}
\label{fig:3}
\end{figure}

Now we discuss the competition of an adaptive agent (\ref{ada}) against an agent with a time-independent efficiency $\eta$. Appendix \ref{ap_d} studies this situation in detail. Here we summarize the main results.

\begin{figure}[t]
\includegraphics[width=7cm]{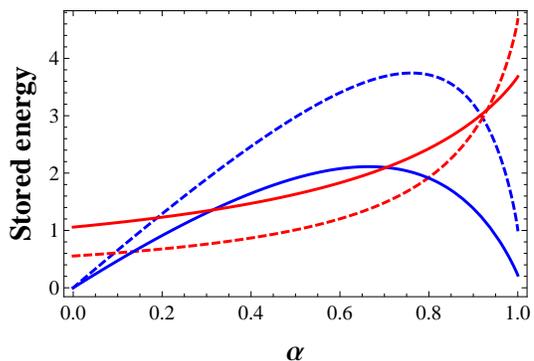}
\caption{Competition between adaptive agent with parameter $\a$ in (\ref{ada}) and a non-adaptive agent with a fixed efficiency $\eta$. The figure shows stored energies|blue and red for (resp.) adaptive and non-adaptive agents|versus $\a$ for a fixed $\eta$. Full (dashed) curves refer to $\eta=\frac{1-\vartheta_{0}}{2}$ ($\eta=1-\sqrt{\vartheta_{0}}$), where the initial temperature ratio $\vartheta_{0}=0.1$ and $\mu=1$ in (\ref{urd44}). 
}
\label{fig:11}
\end{figure}

The adaptive engine operates till the full depletion of the resource. The non-adaptive agent will operate during a finite time $\tau$ (see Appendix \ref{ap_d} for details), i.e. the adaptive agent will still function alone for  $t>\tau$. Fig.~\ref{fig:3} shows the competition results: the adaptive agent wins whenever it competes against sufficiently small or large values of $\eta$. Whenever $\eta$ is around the initial maximal power value $(1-\vartheta_0)/2$ [cf.~(\ref{mp})], also $\a$ should be around the adaptive maximal power regime $\a=1/2$ for adaptive agent to win. In particular, $\a=\frac{1}{2}$ wins against any $\eta$, while $\a\to 1$ (i.e. the optimal $\a$ for a single agent without competition) looses to any $\eta$; see Fig.\ref{fig:3}. However, as Fig.\ref{fig:11} shows, $\a=\frac{1}{2}$ is not the value, where the stored energy of the adaptive agent is maximized. 

Recall that $\eta=1-\sqrt{\vartheta_{0}}$ and $\a \approx 1 $ determine the optimal efficiencies for the single (without competition) non-adaptive and adaptive agent, respectively. Fig.\ref{fig:3} shows that if these values are kept under competition, then the non-adaptive agent wins.

\subsection{Outlook}

ATP stores energy: it is metastable at physiological conditions, and has the half life-time of several hours, after which it hydrolyses spontaneously dissipating the stored energy. Thermodynamically, ATP is similar to a high-temperature bath that stores energy at temperature $T$ \cite{mcclare,jaynes}; cf.~the heat-engine model (\ref{bo}--\ref{mp}).

ATP production in cells is an example of energy extraction and storage that occurs via fixed biochemical pathways \cite{voet}. Two such pathways of ATP production from glucose are well-known: fermentative (aerobic and anaerobic) and oxidative respiration. There is a form of power-efficiency dilemma here: oxidation pathway results to 18 times greater ATP molecules than by fermentation path, while the fermentation pathway is faster (100 times in muscle cells) and hence is more powerful \cite{voet,melkon,shuster1}. 
These pathways have been extensively studied in competitive environments, both in yeasts \cite{shuster,shuster1,maclean,veiga,aledo1,aledo2}  and in solid tumor cells \cite{zheng,vander,liberti,hanahan}. 

We suggest that ATP production paths refer to agents with different internal structure that eventually determines the difference in power and the overall stored energy. Our results in section \ref{arno} mean that competition favors lower efficiencies. This is indeed observed in yeasts and tumor cells \cite{shuster,shuster1,maclean,veiga,aledo1,aledo2}  and also in solid tumor cells \cite{zheng,vander,liberti,hanahan}. Whether also higher powers are favored in our model is a more convoluted question. It does have a straightforward positive answer for the adaptive situation, where the maximal power regime is well-defined at all times.

Note that whenever the resources are shared under a centralized control|i.e. the competion is eliminated and we effectively have a single agent|the optimal efficiency is higher than the efficiency at the maximum power; see (\ref{ca}) and (\ref{mp}). Here the optimal efficiency is defined as the maximizer of the total extracted work. This finding can explain why the oxidative respiration is effectively ubiquitous in multicellular organisms, whereas when the control is lost (e.g. in tumors) cells may switch to fermentative ATP production.

Ref.~\cite{shuster} developed a dynamic model for several populations feeding on a common infinite (i.e. constantly renewable) source. The resource extracted by a population is equalized to its growth per capita. While the model does not hold the laws of thermodynamics (e.g. nothing prevents larger than one efficiencies), it demonstrates the exclusion rule: only the population with the largest power of extraction survives. The size of the survived population is determined by the efficiency of resource extraction. But the proper power-efficiency trade-off is absent [cf.~section \ref{II}], the populations anyhow strive for largest efficiency. It was proposed that this exclusion rule can apply to evolution of ATP production pathways and explain the emergence of multi-cellular organisms: alike cells gather together, thereby exclude competitors and develop less powerful, but more efficient mechanisms of resource extraction \cite{shuster}. Some results of Ref.~\cite{shuster}, e.g. on the relevance of power during the competition, broadly agree with our analysis. However, since our model explicitly agrees with thermodynamics, it provides a richer perspective, e.g. the power uniquely determines the competition outcome only for the adaptive situation. Also, the strict exclusion rule need not hold. 

The standard prisoner's dilemma game was also discussed in the context of agents operating by different ATP production paths \cite{shuster1,aledo2}. In our situation, an effective sub-game that resembles (but is more complicated than) the prisoner's dilemma emerges out of the thermodynamic competition between two agents. It is likely that an effective prisoner's dilemma situation is a reduced description for a more general class of models for agents exploiting the same finite source in accordance with the laws of thermodynamics. We shall explore this hypothesis elsewhere.

\comment{
Above presented prisoner's dilemma game--aroused by thermodynamic consideration in the  competition of non-adaptive agents (agents operating by different ATP production paths), has been discussed in the context of population dynamics \cite{shuster,shuster1}. There the ATP production is related to the reproduction of the given type of agent, as such competition favors high rate/low yield strains (less efficient types in the context of our discussion). ATP production in different regime of glycolisis (dissipative and efficient) has been discussed in \cite{aledo2}, where agents operating in different regimes may face $2\times 2$ games. In these games the payoffs are defined by the work extraction and entropy production on the paths.  }

\section{Summary}
\label{V}

The metabolism of living organisms obeys the laws of thermodynamics, while the energy extraction and storage mechanisms are subject to evolutionary pressure \cite{yang}. In this work we studied competition between energy-extracting thermodynamic agents modeled as heat engines. 

Efficiency and power are two main characteristics of heat engines. They are complementary to each other \cite{novikov,curzon,broeck,mahler,armen}. 
This known power-efficiency tradeoff \cite{novikov,curzon,broeck,mahler,armen} is also observed in various biological systems \cite{dill1,dill2,angulo,brown,roach,shuster,aledo1,aledo2,hans,spitz,tess}. The biological situation is however fundamentally more complex, since it involves adaptation to the source, various channels of using the extracted energy, their mutual feedback {\it etc}.

Models describing these processes in detail are yet to be developed. Here we treat competing agents as heat engines that extract energy (work) from the available source (high-temperature bath). We focus on the linear thermodynamic regime, where the implementation details of the engine are not essential. Two general scenarios for selection are considered: competition for a fixed energy current (the source is then effectively infinite) and for a depletable (finite) source. In both cases, agents compete indirectly, i.e. the interaction between agents is resource mediated. The optimization targets for these scenarios are (resp.) the power (= energy extracted per time unit) and the stored (= total extracted) energy. Both are relevant biologically. 

Competition for the fixed energy current is considered under two known set-ups of game theory: Stackelberg equilibrium \cite{myerson,st,we1} and Pareto optimality \cite{nasho,myerson,luce}. 
Once photosynthesis is a heat engine operating between a hot thermal bath (photons generated by Sun) and the cold thermal bath (Earth environment), these set-ups are relevant for plants competing for light. We show that
Stackelberg's set-up reproduces features observed in invading plants  \cite{gior,ked,weiner,smith,alpert,mor,ipon,funk,falster,anten,funk2}. Pareto optimality assumes weak cooperative behavior of agents, where the resource allocation is fair and is based on the demands of agents. We implemented the Nash bargaining program \cite{nasho,myerson,luce}, as a solution for the fair allocation problem. We observed that the optimal efficiencies of competing agents are higher than the optimal efficiency of a single agent (without competition). This is in stark contrast to what happens in the competition for a finite (depletable) source; see below. The proposed scenario of fair resource allocation is relevant for multicellular organisms, e.g. the allocation between germline and soma cells \cite{kirk,kirk2,chen,mak}. 

We further examine adaptive and non-adaptive agents competing for depletable resources (finite high-temperature thermal bath).
The difference between the two types of agents is the ability of the former to alter its internal structure adjusting it to changes in the exploitable source. Properties of the non-adaptive agent remain fixed in time. The fermentative and oxidative ATP production in cells, which are among the most fundamental, universal biochemical pathways in all life forms, are examples of energy production processes that proceed along fixed pathways \cite{voet}. When these alternative pathways of ATP production are considered as competing agents, the adaptation ability of agents is directly related to the phenotypic adaptation of organisms  \cite{ham,ham1,ham2,marek,toloz,roach,mey,stearn,fors,pier}. 

The optimal efficiencies of the work-extraction process differ for adaptive and non-adaptive agents. The
adaptive agent wins over other agents if its efficiency is that of the maximal power. This however not the efficiency value that maximizes the stored energy in case of competition with non-adaptive agent. The optimal efficiency for the agent with a fixed structure has non-unique (and richer) optimal values. One reason behind this difference is that adaptive agents compete until the final depletion of the sources, whereas non-adaptive agents are unable to do so. We stress that our treatment of adaptation is to some extent formal, since it so far does not account for its full energy cost \cite{ada}: future models should show how to direct a part of the stored energy to the needs of adaptation \cite{gorban}. 

One general outcome of our model is that agents competing for stored energy face analogues of the classical prisoner's dilemma \cite{shuster,we1,myerson,aledo2,hof}. Here the cooperative behavior refers to efficiencies that are optimal in the absence of competition (i.e. for a single agent). The mutually cooperative behavior is still beneficial under competition, but it is not stable, when one agent employs a smaller efficiency (defection). This puts the agents in the best response cycle, where they store a smaller amount of energy. It is possible that prisoner's dilemma is a general consequence of thermodynamic laws applied to exploitation of a depletable resource. 

Overall, we found that simple models with no special assumption beyond the laws of thermodynamics can recapitulate certain features of biological evolution.

\comment{
Operating at the optimal efficiency level yields less stored energy compared with non-optimal (cooperative) behavior, which corresponds to the efficiency level that is optimal in the absence of competition. Therefore, the cooperative behavior is unstable as in the formulation of the prisoner's dilemma. However, the dilemma does not occur in the competition between different types of agents since operating by its optimal efficiency level is unfavorable for adaptive agents and extremely favorable for non-adaptive ones (see Fig.{\ref{fig:11}}). 

Here, our motivation from biology is based on the observation that in multicellular organisms the resource intake process is well controlled. Thus, we may consider different types of cells as agents with various energy demands (defined by an epsilon) and find the optimal allocation of resources. 

This interpretation is the basis of the known "the expensive germline" hypothesis, an assumption of disposable soma theory.  According to the hypothesis, the maintenance of germ cells is assumed to be more costly than the soma cells. 
Indeed, we observe that the solution to the bargaining problem distributes the resources in a non-even way, such that the one with greater energy demand obtains more. The difference in the obtained resources of high and low-demand cells is increasing with the available resource level. As a consequence, in the lack of resources, an organism might more evenly distribute the resources between germ and somatic cells.

We distinguish adaptive and non-adaptive agents in the competition for the same finite and exhaustible resources. 
The internal structure (difference in energy levels) of non-adaptive agents remains fixed during the time. In the absence of competition, the optimal efficiency level of the non-adaptive agent becomes equal to the efficiency level known from endogenous thermodynamics. 
Therefore, the optimality of efficiency levels dramatically changes in the competition. 
Furthermore, we show that there is no unique optimal efficiency level in the competition. Indeed, competition of two non-adaptive agents results in the two-player asymmetric game, where the strategies are fixed and definite efficiency levels, which are below the optimal efficiency level in the absence of competition. The unique Nash equilibrium of the game is in the mixed strategies, identifying coexistence and cyclic dominance between efficiency levels.
As a biological motivation, we discuss the ATP production pathways in the cells. Oxidative respiration can be associated with the heat engine operating by optimal efficiency in the absence of competition. Indeed, oxidative respiration is characteristic for multicellular organisms, in which the competition between cells is under the control. 
Meanwhile, the ATP production pathways are different in the yeasts and solid tumor cells. In solid tumor cells, the environment is highly competitive, and tumor cells are using oxidative fermentation (fermentation even in the presence of oxygen, thus hypoxia is not a unique reason for pathway changes). 
Oxidative fermentation and glycolysis are less efficient than oxidative respiration. Thus, the existence of both fermentative pathways justifies our result that there is no unique optimal efficiency level in the competition of non-adaptive agents.

Adaptive agents can change their internal structure in contrast to non-adaptive agents. Due to these changes the efficiency level of the adaptive agents remains on the same level compared to the maximum possible efficiency level at any instant of time. 
In the absence of competition, the optimal efficiency level tends to be as close to Carnout's efficiency level as possible but not equal. The result is reasonable since efficiency levels close to Carnout's value minimizes the losses in the energy storage process, as a drawback operates at low power.  
Competition of adaptive agents results in the unique optimal efficiency level, the efficiency level of maximum power regime at any time. Competition of adaptive engines lasts till the final depletion of available resources, i.e. an agent with larger instantaneous stored energy outcompetes its opponent. 

The adaptation mechanism is similar to the phenotypic plasticity of different species when organisms are changing their internal structure to meet their metabolic requirements. The competition of adaptive agents is also in a close link with niche construction theory. However, in discussed scenario adaptation of agents is myopic, it follows the environmental changes and is not based on the perception about the future states of the environment. 

Therefore, we show that the maximization of power is not a universally optimal strategy for adaptive agents by considering the competition of adaptive and non-adaptive agents. Adaptive agents win the competition for intermediate efficiency levels. Indeed, the adaptive agent operating at maximum power regime is unbeatable against any non-adaptive agent. However, this regime is not the optimal one in the sense of stored energy during the competition. The winning regions of efficiency levels of adaptive agents are greater for the scare initial resources than for the rich initial resources.

Overall, we show that the competition of the simplest heat engines can describe too different phenomenons observed in different organizational levels. 

}

\acknowledgements This research was supported by the Intramural Research Program of the National Library of Medicine at the NIH.
A.E.A was supported by SCS of Armenia, grants No. 21AG-1C038 and No. 20TTAT-QTa003. 
A.E.A. was partially supported by a research grant from the Yervant Terzian Armenian National Science and Education Fund (ANSEF) based in New York, USA. We thank the late Guenter Mahler for discussions on thermodynamics of evolution.

\onecolumngrid

\appendix

\section{Markov model for heat engine: Structure, Power and Efficiency.}
\label{ap_a}

Here we use slightly different notations as compared to the main text. The notations are related as follows, $\rho_{i\leftarrow j} \equiv \rho_{ij}$, and
\begin{eqnarray}
&& T \equiv T_{21},~~T_{h} \equiv T_{31}, ~~T_{c}\equiv T_{32},\nonumber\\
&& \beta \equiv \beta_{21},~~\beta_{h} \equiv \beta_{31}, ~~\beta_{\rm c}\equiv \beta_{32},\nonumber
\\&& J \equiv J_{21},~~ J_{h} \equiv J_{31},~~J_{c} \equiv J_{32},\nonumber
\end{eqnarray}

Let us recall the simplest model of heat engine \cite{ada}. The model has three states $i=1,2,3$, which is the minimal number of states a stationary operating heat engine can have, because it ought to be in a non-equilibrium state (i.e. to support one cyclic motion), and because it has to support three external objects: one work-source and two thermal baths. Each state $i$ has energy
$E_i$. Transitions between different states are caused by thermal baths that can provide or accept necessary energies. We assume that the resulting dynamics is described by a Markov master equation
\begin{eqnarray}
\label{1}
\dot{p}_i\equiv \frac{d p_i}{d t}= {\sum}_{j}[\rho_{i
  j}p_j-\rho_{j i}p_i], \quad i,j=1,2,3,
\end{eqnarray}
where $p_i$ is the probability to find the system in state $i$ at time $t$, and where $\rho_{ij}$ is the transition rate from state $j$ to state $i$. 

We assume that each pair of transitions $i\leftarrow j$ and $j \leftarrow i$ is caused by thermal baths $T_{ij}=T_{ji}=1/\beta_{ji}$, which are in thermal equilibrium (this point will be clarified latter, when we discuss temperature changing process). The equilibrium nature of thermal baths impose detailed balance condition on the transitions \cite{kampen}
\begin{eqnarray}
  \label{eq:1w}
  \rho_{i j}\, e^{-\beta_{ij}E_j}=  \rho_{j i}\, e^{-\beta_{ij}E_i}, \qquad
\beta_{ij}=\beta_{ji}.
\end{eqnarray}
We take one temperature infinite \cite{ada}: $\beta_{21}=0$. This bath is then a
work-source. This important point is explained via the following related arguments.
First, note that an infinite temperature thermal bath exchanges energy without 
altering its own entropy, which is a feature of mechanical device (i.e. sources of work) \cite{ada}.
Indeed, due to the equilibrium thermodynamic relation $d S_{21}=\beta_{21} d Q_{21}=0$ it
exchanges energy $d Q_{21}\not=0$ at zero entropy change $d S_{21}=0$. Second, if an infinite temperature thermal bath interacts to any (positive) temperature bath, then the former bath always looses energy indicating on no additional costs for the transfer of energy that was stored at an infinite temperature. In that sense, the latter energy is freely convertible to any form of heat, as expected from work.

For the thermal bath with temperature $T_{ij}$, we define $J_{ij}$ as the average energy lost by the bath (for $J_{ij}>0$) or gain ($J_{ij}<0$) per unit of time. Since each bath causes only one pair of transition we get
\begin{eqnarray}
  \label{eq:2w}
J_{ij}=J_{ji}=(E_i-E_j)(\rho_{i j}p_j-\rho_{j i}p_i),
\end{eqnarray}
In the stationary (but generally non-equilibrium) state the average energy of the three-levels system is constant: 
\begin{eqnarray}
\label{fr}
\sum_{i=1}^3\dot{p}_iE_i=0,
\end{eqnarray}
and hence we get from (\ref{eq:1w}, \ref{eq:2w}, \ref{fr}) that 
the sum of energy current nullifies:
\begin{eqnarray}
\label{kr}
J_{12}+J_{23}+J_{13}=0,
\end{eqnarray}
which is the first law of thermodynamics in the stationary state \cite{callen}. 

The stationary probabilities $p_i$ are found from (\ref{1}):
\begin{eqnarray}
  \label{eq:3w}
  p_i=\frac{1}{{\cal Z}}
  [\rho_{i j}\rho_{i k}+\rho_{i
    j}\rho_{j k}+\rho_{i k}\rho_{k j}], 
\end{eqnarray}
where $i\neq j\neq k$, $i,j,k=1,2,3$.

Using (\ref{eq:1w},\ref{eq:2w},\ref{eq:3w}) and noting that $\rho_{1 2}=\rho_{2 1}$ due to the assumed condition $\beta_{12}=\beta_{21}=0$, we obtain for the average energy currents
\begin{eqnarray}
  \label{eq:4w} 
  J_{21}=\frac{\E_2}{{\cal Z}}\, \rho_{2 1}\,\rho_{1 3}\,\rho_{3 2}\,\left[ 1-e^{ (\beta_{32}-
\beta_{31})\E_3-\beta_{32}\E_2 } \right],\\
\label{eq:5w}
J_{31}=-{\E_3J_{21}}/{\E_2},\qquad 
J_{32}={(\E_3-\E_2)J_{21}}/{\E_2},\\
\E_2\equiv E_2-E_1, \qquad \qquad \E_3\equiv E_3-E_1.
\label{bela}
\end{eqnarray}
The heat engine functioning is defined as
\begin{eqnarray}
  \label{eq:6w}
0>  J_{21}=-(E_2-E_1)(p_2-p_1)\rho_{1 2},
\end{eqnarray}
i.e. the infinite-temperature bath gains energy. 
Eq.~(\ref{eq:4w}) implies that for the heat engine functioning it is necessary that
\begin{eqnarray}
  \label{eq:7w}
\E_2[ (1-\vartheta)\E_3-\E_2]>0, \quad 
\vartheta\equiv\beta_{31}/\beta_{32}.  
\end{eqnarray}

We will assume that $\vartheta<1$, that is to say $T_{31}$ ($T_{32}$) is the temperature of the hot (cold) bath. Eq.(\ref{eq:7w}) shows that the system will operate as a heat engine for a given $\vartheta<1$ if $\E_3>\E_2$.

The efficiency of any heat engine is defined as the result (i.e. the
extracted work) divided over the resource (i.e. the energy coming from the hot bath).
Under $\vartheta<1$ and $\E_3>\E_2$ the efficiency $\eta$ amounts to 
\begin{eqnarray}
  \label{eq:000}
  \eta\equiv\frac{-J_{21}}{J_{31}} =\frac{\E_{2}}{\E_{3}}
  \,\leq\, \eta_{\rm C}\equiv
  1-\vartheta,
\end{eqnarray}
i.e. the efficiency is bounded from the above by the Carnot efficiency
$\eta_{\rm C}$. Hence, (\ref{eq:000}) is the general message of the second law for the heat engine efficiency \cite{callen}.

Another important message of (\ref{eq:000},\ref{eq:7w},\ref{eq:4w}) is the power-efficiency trade-off: at the maximal efficiency the power $-J_{12}$ of the heat engine nullifies. This trade-off is also a general feature of heat engines \cite{mahler}, though it is frequently missed in phenomenological treatments of equilibrium thermodynamics \cite{callen}. This trade-off is one pertinent reason for having an explicit microscopic model of a heat engine \cite{mahler}. 

\subsubsection{High temperature regime.}

The work power $J_{12}$ in (\ref{eq:4w}) depends on the specific form of the transition rates  
$\rho_{i j}$ that enter the detailed balance condition (\ref{eq:1w}). The form of $\rho_{i j}$ depends on the physical implementation of the model; e.g. the Arrhenius form of $\rho_{i j}$ applies in chemical reaction dynamics \cite{kampen}. We  shall work in the high-temperature limit, where the details of $\rho_{i j}$ are not important provided that it holds the detailed balance (\ref{eq:1w}). Now $\beta_{32}\ll 1$ and $\beta_{31}\ll 1$ are sufficiently small, i.e. $E_i\beta_{32}\ll 1$ and $E_i\beta_{31}\ll 1$, but $0\leq \vartheta\leq 1$ in (\ref{eq:7w}) can be arbitrary. In this limit the power of heat engine reads from (\ref{eq:4w}):
\begin{eqnarray}
  \label{eq:1b}
  J_{21}= -\rho\beta_{32} \E_{2}((1-\vartheta)\E_{3}-\E_{2}),
\end{eqnarray}
where we denoted $\rho=\frac{1}{{\cal Z}}\, (\rho_{2 1}\rho_{1 3}\rho_{3 2})|_{\beta_{31}=\beta_{32}=0}$.
We shall assume that $\rho$ is a constant. Note that for $\beta_{31}=\beta_{32}=0$ we get $\rho_{ij}=\rho_{ji}$ from (\ref{eq:1w}). Hence in this limit we get $p_i=1/3$; cf.~(\ref{eq:3w}). Now $J_{21}\not =0$ in (\ref{eq:1b}) due to small deviations of $p_i$ from $1/3$ that exist due to a small but finite $\beta_{31}$ and $\beta_{32}=0$.

In the high-temperature regime our model is fully consistent with linear equilibrium thermodynamics \cite{caplan,broeck}. Structurally similar models were widely employed in bioenergetics for modeling coupled chemical reactions, where the passage of heat from higher to lower temperatures corresponds to the down-hill reaction, where the work-extraction refers to the up-hill reaction \cite{caplan}. However, we emphasize that a microscopic model clarifies the status of the involved parameters, as well as demonstrates explicitly that the extracted work relates to the stored energy. These two important aspects are not clear within the phenomenological introduction of linear thermodynamics models. 


\section{Competition for a fixed energy current}
\label{ap_b}
  
Two agents competing for fixed resources can be described as 
two heat engines [cf.~(\ref{eq:1b})]:
\begin{eqnarray}
  \label{ora}
  J^{[k]}_{21}&=& -\rho\beta_{32} \E^{[k]}_{2}((1-\vartheta)\E^{[k]}_{3}-\E^{[k]}_{2})\\
  &=&
  -\rho\beta_{32} \E^{[k]\, 2}_{3}\eta_k(1-\vartheta-\eta_k), \quad k=1,2,
\end{eqnarray}
where $\eta_k=\E^{[k]}_{2}/\E^{[k]}_{3}$ are the efficiencies [cf.~(\ref{n4})]. 
Note that for the photosynthesis $\theta\simeq 10^{-3}$ is a small number.

We write this constraint (\ref{fix:1})as
\begin{eqnarray}
  \label{ora2}
\frac{  |J^{[1]}_{21}| }{\eta_1}+\frac{  |J^{[2]}_{21}| }{\eta_2}\leq A', \quad {\rm or},\\
\E^{[1]\, 2}_{3} (1-\vartheta-\eta_1)+\E^{[2]\, 2}_{3} (1-\vartheta-\eta_2)\leq A'',
  \label{ora3}
\end{eqnarray}
where $A'$ and $A''$are positive constants.

\subsubsection{Game-theoretical approach: actions and utilities}

\BEA
\eta_1+\epsilon\eta_2\geq A, \qquad \epsilon\equiv (\E^{[2]}_{3}/\E^{[1]}_{3})^2. 
  \label{ra4}
\EEA
To simplify the parametrization of the problem we denote
\BEA
x_k=\eta_k/(1-\vartheta),\quad k=1,2,
\EEA
and write the extracted works as
\BEA
&& - J^{[k]}_{21}=\rho\beta_{32} \E^{[1]\,2}_{3}(1-\vartheta)^2{\cal W}_k, \quad k=1,2,\\
 {\cal W}_1&=&x_1(1-x_1),\quad {\cal W}_2=\epsilon x_2(1-x_2).
  \label{ra5}
\EEA
The problem described by (\ref{ra4}--\ref{ra5}) has a game-theoretic interpretation, where 
${\cal W}_1$ and ${\cal W}_2$ refer to utilities of each agent, and (\ref{ra4}) ensures that their actions $x_1$ and $x_2$ are coupled. Hence, for $A$ in (\ref{ra4}) we shall assume 
\BEA
\frac{(1+\epsilon)(1-\vartheta)}{2}<A<{(1+\epsilon)(1-\vartheta)}.
\label{ra6}
\EEA
The first inequality here means that the agent cannot simultaneously maximize ${\cal W}_1$ and ${\cal W}_2$ at 
$\eta_1=\eta_2=(1-\vartheta)/2$. The second inequality means that the constraint (\ref{ra4}) still allows some  
$\eta_1<1-\vartheta$ and $\eta_2<1-\vartheta$. Introducing a new constant via $A=a(1+\epsilon)(1-\vartheta)$ we conveniently write (\ref{ra4}) as
\BEA
x_1+\epsilon x_2\geq a(1+\epsilon),\qquad 1\geq a\geq 1/2,
\label{ra77}
\EEA
where the last inequalities follow from (\ref{ra6}).

\subsubsection{Pareto line and the bargaining equilibrium}

Pareto's concept of equilibrium assumes a certain level of passive cooperation between ${1}$ and ${2}$. Now ${1}$ does not increase its utility ${\cal W}_1$ without decreasing the utility ${\cal W}_2$ of ${2}$, and likewise ${2}$ does not increase ${\cal W}_2$ without decreasing ${\cal W}_1$ \cite{nasho,myerson,luce}. This contrasts to Stackelberg's solution, where the first moving agent chooses the optimal condition for itself, irrespective of what happens to another agent. Not surprisingly, Pareto's concept does not provide any unique choice of $(x_1, x_2)$, and an additional reasoning is needed for selecting a unique outcome among Pareto solutions. 

It is seen from (\ref{ra5}, \ref{ra77}) that the set of Pareto equilibria (i.e. the Pareto line) is given by the following three conditions
\BEA
\label{nana}
x_1+\epsilon x_2= a(1+\epsilon), \quad 0\leq x_k\leq 1, \quad k=1,2, \\ 
x_2 \in \left[{\rm max}[\frac{1}{2},\,\,\frac{1}{\epsilon}(a(1+\epsilon)-1)], \right.\nonumber \\ \left.{\rm min}[\frac{1}{\epsilon}(a(1+\epsilon)-1),\,\,1]
\right]\nonumber. 
\EEA
Note that there is a continuum of Pareto equilibria. Moreover, for this game all Pareto equilibria are also Nash equilibria. Indeed, recall that the Nash equilibrium looks for a pair $({x}^*_1(x_2), {x}^*_2(x_1))$ such that ${x}^*_1(x_2)$ is the best response to ${x}_2$ (i.e. the conditional maximum of ${\cal W}_1$ given ${x}_2$), while ${x}^*_2(x_1)$ is the best response to ${x}_1$ \cite{nasho,myerson,luce}.

 Nash's bargaining solution is one approach for making a unique choice within (\ref{nana}) \cite{nasho,roth,myerson,luce}. For defining the bargaining target we need to find the worst utility for each agents: 
 \BEA
 {\cal W}^*_1={\rm max}_{x_1}{\rm min}_{x_2} [\, {\cal W}_1\, ],\,\,\,
 {\cal W}^*_2={\rm max}_{x_2}{\rm min}_{x_1} [\, {\cal W}_2\, ].
 \EEA
 To find out $ {\cal W}^*_1$, note from (\ref{ra77}) that any choice $x_1$ of ${1}$ should hold $x_1\geq a(1+\epsilon)-\epsilon x_2$. If ${2}$ arranged its action $x_2$ such that $a(1+\epsilon)-\epsilon x_2\geq 1$, then there is no choice to be made by ${1}$. For such cases we shall define its utility ${\cal W}_1=0$, and this is clearly the worst outcome for ${1}$. Now $a(1+\epsilon)-\epsilon x_2\geq 1$ or equivalently $[a(1+\epsilon)-1]/\epsilon\geq x_2$ will work provided that $a(1+\epsilon)-1\geq 0$. Hence, if the latter condition holds we get ${\cal W}_1=0$. Otherwise, if $a(1+\epsilon)-1< 0$, then ${2}$ achieves ${\rm min}_{x_2} [\, {\cal W}_1\, ]$ when the bound $x_1\geq a(1+\epsilon)-\epsilon x_2$ is possibly tight, i.e. for $x_2=0$, and then $x_1= a(1+\epsilon)={\rm argmax}_{x_1}{\rm min}_{x_2} [\, {\cal W}_1\, ]$. Altogether, we get 
\BEA
\label{ded1}
{\cal W}^*_1=a(1+\epsilon)\,{\rm max}[0,1-a(1+\epsilon)].
\EEA
${\cal W}^*_2$ is calculated in the same way:
\BEA
{\cal W}^*_2=a(1+\epsilon^{-1})\,{\rm max}[0,\epsilon-a(1+\epsilon)].
\label{ded2}
\EEA
Note that only one among ${\cal W}^*_{1}$ and ${\cal W}^*_{2}$ can be non-zero. Indeed, from (\ref{ded1}) and (\ref{ded2}) it follows that ${\cal W}^*_{1},\, {\cal W}^*_{2}>0$ if $\frac{a}{1-a}<\epsilon<\frac{1-a}{a}$ which is impossible since $1\geq a\geq \frac{1}{2}$. The fact of e.g. ${\cal W}^*_{1}>0$ is important, since it means that ${1}$ cannot be eliminated from the competition. Now 
(\ref{ded1}, \ref{ded2}) are not simply the separate global minima of $({\cal W}_{1},{\cal W}_{2})$, i.e. generally $({\cal W}^*_{1},{\cal W}^*_{2})\not=(0,0)$, e.g. because generally ${\cal W}^*_{1}=0$ is not reached for any action of ${2}$. (Still we get $({\cal W}^*_{1},{\cal W}^*_{2})=(0,0)$ for $\epsilon=1$.)

Now the Nash bargaining solution $(\hat x_1,\hat x_2)$ is found from maximizing the geometric mean of ${\cal W}_{1}-{\cal W}^*_{1}$ and ${\cal W}_{2}-{\cal W}^*_{2}$ \cite{we2}:
\BEA
{\rm max}_{x_1,x_2}\left[ \, ({\cal W}_{1}(x_1)-{\cal W}^*_{1})({\cal W}_{2}(x_2)-{\cal W}^*_{2})\,\right],
\label{raa}
\EEA
where $(x_1, x_2)$ should vary along all allowed values defined by $x_1+\epsilon x_2\geq a(1+\epsilon), \quad 0\leq x_k\leq 1,$ and by the values corresponding the worst outcomes of the players ${\cal W}^*_{k}$.  As it is mentioned above, it is possible that in the worst case $1$ will not have a choice for the given action of $2$. That is, the action of $1$ will be lie outside of the allowed region. Thus, we obtain the following region of maximization for  (\ref{raa}), where the worst outcomes are also take an account.

\BEA
{\rm max}[0, (a(1+\epsilon)-1)/\epsilon]<x_2<{\rm min}[1, a(1+\epsilon^{-1})].
\EEA

Note, that the last bounded region includes the Pareto line of the problem (\ref{nana}). One may even maximize (\ref{raa}) over the Pareto line, since the  outcome of the maximization will lie on that line \cite{we2, roth}.

Eq.~(\ref{raa}) maximizes a mean, since ${1}$ and ${2}$ are assumed to be equivalent, and it is the geometric mean, since generally dimensions of utilities ${\cal W}_{1}$ and ${\cal W}_{2}$ can be different. Ref.~\cite{we2} discusses axioms of the bargaining solution and shows that they are consistent with thermodynamics. 

\section{Strategies of work-extraction. Depletable resources.}
\label{ap_c}

The dynamics of $T_{31}$ can be deduced from the formula of equilibrium heat-capacity,
which governs the change of temperature for an equilibrium body given the change of its internal energy \cite{callen}. Once the changes of $T_{31}$ are assumed to be slow:
\begin{eqnarray}
\label{eq:1ca}
\frac{d T_{31}}{d t}=-C^{-1}J_{31}
\end{eqnarray}
where $C=C_{31}$ is the heat capacity of thermal baths. Indeed, $C$ is large, since it scales with the number of the bath degrees of freedom. On the other hand, $J_{31}\sim 1$, hence $\frac{d T_{31}}{d t}$ is small, i.e. the change of $T_{31}$ is slow. We shall make a natural assumption that it is much slower than the relaxation of the engine to its stationary state (\ref{eq:3w}).

Taking into an account that $\beta_{32}={\mit const}$, we find for $\vartheta=\beta_{31}/\beta_{32}$
from (\ref{eq:1b}, \ref{eq:5w}):
\begin{eqnarray}
\label{eq:2c}
\frac{d \vartheta}{d t}= C^{-1}\beta_{32}^2\rho \vartheta^2\,
\E_{3}((1-\vartheta)\E_{3}-\E_{2}).
\end{eqnarray}

Different strategies of work-extraction refer to differences in its structural parameters $\E_2$ and $\E_3$.

\subsubsection{Non-adaptive agent}.

Here we discuss the case, when the heat engine is adapted to the initial temperature ratio, i.e the efficiency $\eta$ of the heat engine remains fixed in the course of time.   
Eq.~(\ref{eq:1b}) shows that for a fixed $\vartheta$ the maximum power of heat engine (i.e. the maximum of $|J_{21}|$) is attained for 
\begin{eqnarray}
  \label{eq:2b}
  \frac{\E_{2}}{\E_{3}}=\frac{1-\vartheta}{2}.
\end{eqnarray}
Eq.~(\ref{eq:2b}) implies that the maximal (Carnot) efficiency is attained for 
$\frac{\E_{2}}{\E_{3}}=1-\vartheta$.
Note that in various comparisons between different engine structures it is sensible to keep the difference  $\E_3=E_3-E_1$ fixed and vary only $\E_2$. Indeed, in the considered regime $\E_3>\E_2>0$, $\E_3$ is the difference between the maximal and minimal energies of the heat engine. Hence fixing $\E_3$ means to fix the global energy scale.

We introduce dimensionless parameters and write (\ref{eq:1b}, \ref{eq:2c}) as

\begin{eqnarray}
\label{urd1}
&&\hat J_{21}=-\eta(1-\vartheta-\eta), ~~ \hat J_{21}= J_{21}/(\rho\beta_{32}\E_3^2),\\
&&\dot\vartheta=\mu\vartheta^2(1-\vartheta-\eta), ~~ \mu=
C^{-1}\beta_{32}^2\rho \E_{3}^2.
\label{urd2}
\end{eqnarray}

For a constant efficiency of the heat engine $\eta$, which includes the cases where the engine is maximizes $|J_{12}|$ for the initial resource,
(\ref{urd2}) is solved as

\begin{eqnarray}
\vartheta(t)=\frac{1-\eta}{1+W\left[\left(\frac{1-\eta}{\vartheta_0}-1\right)\, e^{\frac{1-\eta}{\vartheta_0}-1-\mu(1-\eta)^2 t}  \right]},
\label{lamb}
\end{eqnarray}
where $\vartheta_0=\vartheta(0)$ is the initial value, and $W[z]$ is Lambert's function (or product logarithm), which solves equation $z=We^W$.
Note that $W[z]\to z$ for $z\to 0$, which determines the long-time behaviour 
$\vartheta(t)\to(1-\eta)$ of (\ref{lamb}), in this limit the power of of the heat engine nullifies since the efficiency of the heat engines becomes equal to the Carnot level $\eta=\eta_{\rm C}=1-\vartheta(t)$.

The full dimensionless stored energy is found from (\ref{urd1}) and (\ref{lamb}) as
The full dimensionless stored energy is found from (\ref{urd1}) as
\begin{eqnarray}
&&-\int_0^\infty {\rm d}t\, \hat J_{21}(t)=\eta(1-\eta)\int_0^\infty {\rm d}t\,
\left[1- \left(1+W\left[\left(\frac{1-\eta}{\vartheta_0}-1\right)\, 
e^{\frac{1-\eta}{\vartheta_0}-1-\mu(1-\eta)^2 t}  \right]\right)^{-1}  \right]\nonumber\\
&&=\frac{\eta}{\mu(1-\eta)}\int_0^1 \frac{{\rm d}\hat t}{\hat t}\,
\left[1- \left(1+W\left[\left(\frac{1-\eta}{\vartheta_0}-1\right)\, 
e^{\frac{1-\eta}{\vartheta_0}-1}\,\hat t  \right]\right)^{-1}\right]\nonumber\\
\label{ue2}
&&=\frac{\eta}{\mu(1-\eta)} W\left[\left(\frac{1-\eta}{\vartheta_0}-1\right)\, 
e^{\frac{1-\eta}{\vartheta_0}-1}\right],\\
&&= \frac{\eta}{\mu(1-\eta)}\left(\frac{1-\eta}{\vartheta_0}-1 \right),
\label{uea}
\end{eqnarray}
where (\ref{ue2}) follows from 
\begin{eqnarray}
W'[z]=W[z]/(z(1+W[z])), 
\label{lambert_1}
\end{eqnarray}

and (\ref{uea}) holds due to the definition of $W(z)$. When maximizing (\ref{uea}) over $\eta$ for a fixed $\vartheta_0$ (i.e. for a fixed initial resource), we find that (\ref{uea}) is maximized for $\eta=1-\sqrt{\vartheta_0}$, which is larger than the value $\eta=(1-\vartheta_0)/2$, at which the power $|J_{12}|$ is maximized. Hence, for sufficiently sizable resources it pays to consume them with the efficiency larger than at the maximum of $|J_{12}|$.

 \subsubsection{Adaptive agent}. 
 
 Next, let us consider the case of adaptive agent, where 
\begin{eqnarray}
\label{fox}
\eta(t)=\alpha \eta_{\rm C}(t)=\alpha(1-\vartheta(t)),
\end{eqnarray}
where $0<\alpha\leq 1$. 
This involves situations, where the engine is sufficiently complex to monitor the environment and adapt to time-dependent temperatures, i.e. (\ref{fox}) involves the cases, where the engine maximizes $|J_{12}|$ for
any time-dependent $\vartheta(t)$.
Now (\ref{urd2}) is solved as
\begin{eqnarray}
\vartheta(t)=\left(1+W\left[\left(\frac{1}{\vartheta_0}-1\right)\, 
e^{\frac{1}{\vartheta_0}-1-\mu(1-\a) t}  \right]\right)^{-1}.
\label{lamb2}
\end{eqnarray}

Using the same method as for (\ref{uea}), we get from (\ref{fox}, \ref{lamb2}):

\begin{eqnarray}
-\int_0^\infty {\rm d}t\, \hat J_{21}(t)
&=&\frac{\a}{\mu}\left( W\left[\left(\frac{1}{\vartheta_0}-1\right)\, 
e^{\frac{1}{\vartheta_0}-1}\right]-\ln\left\{  1+
W\left[\left(\frac{1}{\vartheta_0}-1\right)\, 
e^{\frac{1}{\vartheta_0}-1}\right]
\right\} \right),\\
&=&\frac{\a}{\mu}\left(\frac{1}{\vartheta_0}-1+\ln\vartheta_0  \right)
\label{ueda}
\end{eqnarray}

which makes clear that the maximum of (\ref{ueda}) is reached for $\a\to 1$, i.e. once the adaptation (\ref{fox}) is allowed, the maximal stored energy is reached under the most effective scenario of work-extraction. Note that reaching the $\a=1$ limit of (\ref{ueda}) demands an infinite time, since $|J_{12}|\to 0$ for $\a\to 1$.

It is worth noting, that the advanced capability of adaptation is not always provide an advantage over the simpler scenarios of adaptation when the heat engine is operating alone, i.e. without competition. Indeed, comparing (\ref{ueda}) and (\ref{uea}) in the case of the possible maximum efficiency $\eta=1-\sqrt{\vartheta_{0}}$ of the heat engine performing by fixed internal structure for the given initial resources $\vartheta_{0}$ it becomes obvious that the stored energy of the heat engine operating by local adaptation scenario is greater than that of performing by fixed internal structure scenario when 

\begin{eqnarray}
\label{comp}
\alpha \geq \frac{1-2\sqrt{\vartheta_{0}}+\vartheta_{0}}{1-\vartheta_{0}+\vartheta_{0} \ln{\vartheta_{0}}}
\end{eqnarray}

Obviously, the right hand-side of (\ref{comp}) is always $\leq 1$. Thus, the stored energy of the heat engine operating by the local adaptation scenario is greater than that of operating by the fixed internal structure for $\alpha \to 1$. Note that in the case of rich resources $\vartheta \approx 0$ the stored energies of both heat engines operating on the maximum possible efficiencies will be the same.

\section{Two agents (heat engines) on the same resource}
\label{ap_d}
\subsubsection{Competition of two non-adaptive agents.}
Let us now consider two agents (1 and 2) recalling that they are taken to have the same value for $\E_3$ and $\rho$ and do interact with the same thermal baths. Instead of (\ref{urd1}, \ref{urd2}) we shall now have
($k=1,2$)
\begin{eqnarray}
\label{urd3}
&&\hat J^{[k]}_{21}=-\eta_k(1-\vartheta-\eta_k),\qquad \eta_k=\E^{[k]}_2/\E_3,   \\
&&\dot\vartheta=2\mu\vartheta^2(1-\vartheta-\bar \eta),\qquad \bar \eta=(\eta_1+\eta_2)/2.
\label{urd4}
\end{eqnarray}

According to (\ref{urd4}), $\vartheta$ will relax two times faster [than in (\ref{urd2})] to the rest point 
$1-\bar \eta$. If we assume $\eta_1<\eta_2$ than the second agent will need to have a finite life-time determined via $\eta_2=1-\vartheta(\tau_2)$. Otherwise for $t>\tau_2$ this agent will not anymore function as a heat engine. Hence the competition for the same resource brings in a finite life-time of functioning as heat engine.

By analogy to (\ref{lamb}) we solve (\ref{urd4}) as
\begin{eqnarray}
\vartheta(t)=(1-\bar \eta)\left(1+W\left[\left(\frac{1-\bar \eta}{\vartheta_0}-1\right)\, e^{\frac{1-\bar \eta}{\vartheta_0}-1-2\mu(1-\bar \eta)^2 t}  \right]\right)^{-1}.
\label{lamb4a}
\end{eqnarray}
Hence the total extracted work by the second agent is to be calculated from (\ref{urd3}) as
\begin{eqnarray}
-\int_0^{\tau_2} {\rm d}t\, \hat J^{[2]}_{21}(t), \qquad \vartheta(\tau_2)=1-\eta_2,
\label{ued4}
\end{eqnarray}
where $\tau_2$ is the time at which the interaction of the second agent with thermal baths is to be switched off, since for $t>\tau_2$ it will cease to function as a heat engine (i.e. extract work). Noting that $W^{-1}(z)=ze^z$, we find from (\ref{lamb4a},\ref{ued4}):
\begin{eqnarray}
\tau_2=\frac{1}{2\mu(1-\bar \eta)^2}
\ln\left[   
\left(\frac{1-\bar \eta}{\vartheta_0}-1\right)\, 
e^{\frac{1-\bar \eta}{\vartheta_0}-1}\,\,\frac{1- \eta_2}{\eta_2-\bar \eta}\,\, e^{\frac{\bar \eta-\eta_2}{1-\eta_2}}
\right]
\label{koraa}
\end{eqnarray}

Using (\ref{lamb4a}, \ref{ued4}, \ref{koraa}) we obtain:
\begin{eqnarray}
&&-\int_0^{\tau_2} {\rm d}t\, \hat J^{[2]}_{21}(t)=\eta_2(1-\eta_2)\tau_2-\eta_2
\int_0^{\tau_2} {\rm d}t\,\vartheta(t)=
\eta_2(1-\eta_2)\tau_2-\frac{\eta_2}{2\mu(1-\bar \eta)}\ln{\left[\frac{
\frac{1-\bar \eta}{\vartheta_0}-1}{\frac{\eta_2-\bar \eta}{1-\eta_2}}
\right]}.
\label{ued5a}
\end{eqnarray}
Indeed, we have from (\ref{lamb4a}):
\begin{eqnarray}
&&\int_0^{\tau_2} {\rm d}t\,\vartheta(t)=\frac{1}{2\mu(1-\bar \eta)}\int_{e^{-2\mu(1-\bar \eta)^2\tau_2}}^1\frac{{\rm d}\hat t}{\hat t}\left( 1+W\left[   
\left(\frac{1-\bar \eta}{\vartheta_0}-1\right)\, e^{\frac{1-\bar \eta}{\vartheta_0}-1}\,\,
\hat t\right]  
\right)^{-1}\\
&&=\frac{1}{2\mu(1-\bar \eta)}\left(\ln{
W\left[\left(\frac{1-\bar \eta}{\vartheta_0}-1\right)\, e^{\frac{1-\bar \eta}{\vartheta_0}-1}\right]}  
-\ln{W\left[\left(\frac{1-\bar \eta}{\vartheta_0}-1\right)\, e^{\frac{1-\bar \eta}{\vartheta_0}-1}
e^{-2\mu(1-\bar \eta)^2\tau_2}
\right]  }
\right)\\
&&=\frac{1}{2\mu(1-\bar \eta)}\ln{\left[\frac{
\frac{1-\bar \eta}{\vartheta_0}-1}{\frac{\eta_2-\bar \eta}{1-\eta_2}}
\right]}
\end{eqnarray}
The work extracted by the first agent reads
\begin{eqnarray}
-\int_0^{\tau_2} {\rm d}t\, \hat J^{[1]}_{21}(t)-\int_{\tau_2}^\infty {\rm d}t\, \widetilde{\hat J}^{[1]}_{21}(t),
\label{gruma}
\end{eqnarray}
where in the first integral $\vartheta(t)$ is given by (\ref{lamb4a}). In the second integral $\theta(t)$ holds a different equation, 
\begin{eqnarray}
\dot\vartheta=\mu\vartheta^2(1-\vartheta-\eta),
\end{eqnarray}
since for times $t>\tau_2$ only the first agent couples to the baths. 

The first integral in (\ref{gruma}) is calculated analogously to (\ref{ued5a}): 
\begin{eqnarray}
-\int_0^{\tau_2} {\rm d}t\, \hat J^{[1]}_{21}(t)
=\eta_1(1-\eta_1)\tau_2-\frac{\eta_1}{2\mu(1-\bar \eta)}\ln{\left[\frac{
\frac{1-\bar \eta}{\vartheta_0}-1}{\frac{\eta_2-\bar \eta}{1-\eta_2}}
\right]}.
\end{eqnarray}
The second integral in (\ref{gruma}) is
found from (\ref{uea}), where we should change $\eta\to \eta_1$ and $\vartheta_0\to 1-\eta_2$.

Thus the total stored energy of the first agent has the following form

\begin{eqnarray}
\label{grum}
&&-\int_0^{\tau_2} {\rm d}t\, \hat J^{[1]}_{21}(t)-\int_{\tau_2}^\infty {\rm d}t\, \widetilde{\hat J}^{[1]}_{21}(t)= \\
\label{grem}
&&= \eta_1(1-\eta_1)\tau_2-\frac{\eta_1}{2\mu(1-\bar \eta)}\ln{\left[\frac{
\frac{1-\bar \eta}{\vartheta_0}-1}{\frac{\eta_2-\bar \eta}{1-\eta_2}}
\right]}+\nonumber\\
&&+ \frac{\eta_1}{\mu(1-\eta_1)}\bigg(\frac{1-\eta_{1}}{1-\eta_{2}}-1\bigg)\nonumber,
\end{eqnarray}

\subsubsection{Game theoretical analysis. Emergence of $2\times2$ game.}

\begin{figure}[t]
    \includegraphics[width=7cm]{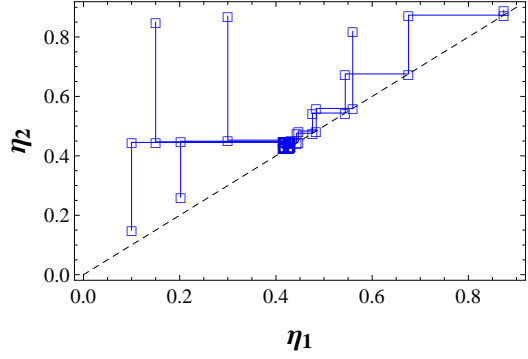}
    \caption{Implementation of best response algorithm for $\vartheta_{0}=0.1$. Each path starts from different initial values $(\eta_1,\eta_2)$ illustrated by rectangles connected to the path by one-side.
    An rectangle on each path represents the steps taken by algorithm. }
    \label{fig:111}
\end{figure}

Let us define the stored energies of each agent (\ref{grum}) and (\ref{ued5a}) by $W_1(\eta_{1},\eta_{2})$ and $W_2(\eta_{1},\eta_{2})$. The stored energies of each agent obviously depends on both efficiency levels $\left\{\eta_{1},\eta_{2}\right\}$. Hence, both agents face game theoretical situation where its obtained payoff depends also on the chosen strategy of its opponent. Here, strategies of each players are the chosen efficiently levels. 

Thus, for each initial temperature ratio having the set of possible strategies $\left\{\eta_{1},\eta_{2}\right\}$--satisfying $\eta_{1}<\eta_{2}\in [0,1-\vartheta_{0}]$, and payoffs $[W_1,W_2]$ defined on the strategy profiles $\left\{\eta_1,\eta_2\right\}$ one may try to find the pure Nash equilibrium of the game. 

Here we employ the best response algorithm for search. The steps of the algorithm are as follows 
\begin{itemize}
    \item initialize the input as arbitrarily point  $\left\{\eta_1,\eta_2\right\}$ such that $\eta_{1}<\eta_{2}$.
    \item Find the maximum of $W_1$ and $W_{2}$ keeping the opponents efficiency level fixed. update the values $\left\{\eta_1,\eta_2\right\}$ from the maximization.
    \item  Repeat until algorithm converges.
\end{itemize}

\begin{eqnarray}
{\rm initial~~input}&& ~~ \left\{\eta_{1},\eta_{2}\right\} \in [0,1-\vartheta_{0}]~ \&\& ~\eta_{1}<\eta_{2},\nonumber\\
{\rm repeat}&&\nonumber\\
    &&\eta^t_{1}={\rm argmax}_{\hat{\eta}} [W_1(\hat{\eta},\eta^t_{2}), ~~0 \leq \hat{\eta}<\eta^t_{2}] \nonumber\\
    &&\eta^t_{2}={\rm argmax}_{\hat{\eta}}[W_2 (\eta^t_{1},\hat{\eta}), ~~\eta^t_{1}<\eta_{2}\leq1-\vartheta_{0}]\nonumber\\
{\rm until}&& \eta^{t+n}_{1}=\eta^t_{1}.\nonumber
\end{eqnarray}

A particular implementation of the algorithm is illustrated in Fig.{\ref{fig:111}} for various initialization $\left\{\eta_1,\eta_2\right\}$ and for $\vartheta_{0}=0.1$.

The algorithm yields a pair of efficiency levels 
$(\eta'_1,\eta''_1)$ and $(\eta'_2,\eta''_2)$ respectively for 1 and 2 agents. The values of the obtained efficiency levels for different initial temperature ratio is given in table\ref{tab1}.

\begin{table}[!htbp]
\begin{tabular}{||c|c|c|c|c|c||} 
 \hline
 $ \vartheta_{0}$ & $1-\sqrt{\vartheta_{0}}$ & $\eta'_{1}$ & $\eta''_{1}$ & $\eta'_{2}$ & $\eta''_{2}$  \\ [1.ex] 
 \hline\hline
 0.1 & 0.687 & 0.416 & 0.427 & 0.447 & 0.427 \\ 
 \hline
 0.25 & 0.5 & 0.279 & 0.293 & 0.315 & 0.293 \\
 \hline
 0.5 & 0.292 & 0.151 & 0.161 & 0.179 & 0.161 \\
 \hline
 0.75 & 0.13 & 0.06 & 0.07 & 0.08 & 0.07 \\
 \hline
 0.9 & 0.05 & 0.024 & 0.026 & 0.03  &0.026 \\ [1.ex] 
 \hline
\end{tabular}
\caption{The values of obtained optimal efficiency levels $ \eta'_1, \eta''_1, \eta'_2, \eta''_2$ and the optimal efficiency level in the absence of competition $\eta=1-\sqrt{\vartheta_{0}}$ for various initial temperature ratio $\vartheta_0$.}
\label{tab1}
\end{table}

As it is seen from the Table.\ref{tab1}, the values of optimal efficiencies slightly differ.

These pairs form a cyclic best response $\left\{\eta'_1,\eta'_2\right\} \to \left\{\eta''_1,\eta'_2\right\} \to \left\{\eta''_1,\eta''_2\right\} \to \left\{\eta'_1,\eta''_2\right\} \to \left\{\eta'_1,\eta'_2\right\}$, i.e there is no mutually preferable outcome for the agents.

Thus, the best response program yields to $2\times2$ game characterized by the presence of cyclic best responses (known as coexistence games). 
Table\ref{tab2} illustrates an example of the resulted $2\times2$ game.

\begin{table}[h!]
\begin{tabular}{||c| c| c||}
     \hline 
     $~$ & $\eta'_2$ & $\eta''_2$ \\ [1.ex]
     \hline
     $\eta'_1$ & $(1.843, 1.726)$ & $(1.770,~1.724)$ \\ [1.ex]
     \hline
     $\eta''_1$ & $(1.846, 1.761)$ & $(1.764, 1.764)$\\ [1.ex]
     \hline\hline
\end{tabular}    
\caption{$2\times 2$ game for $\vartheta_{0}=0.1$. The first/second number in each cell represents the payoffs of the first/second agent .}
\label{tab2}
\end{table}

Now, let us start from the cell $\left\{\eta'_1,\eta'_2\right\}$. We see that for the first player it is better to change its strategy since $W_1\left(\eta'_1,\eta'_2\right)<W_1 \left(\eta''_1,\eta'_2\right)$. Then, the second agent will change its strategy since  $W_2\left(\eta''_1,\eta'_2\right)<W_1 \left(\eta''_1,\eta''_2\right)$. In this way we will recover the whole best response cycle.

It has to be  noted, that in the resulted $2\times2$ game each player receive less than if they would operate by $1-\sqrt{\vartheta_{0}}$, which is the optimal efficiency level in the absence of competition. 

Indeed, the payoff of each player operating by  $1-\sqrt{\vartheta_{0}}$ is obtained from (\ref{uea}) by interchanging $\mu \to 2\mu$. For the considered case $W_1\left(1-\sqrt{\vartheta_{0}},1-\sqrt{\vartheta_{0}}\right) = W_2\left(1-\sqrt{\vartheta_{0}},1-\sqrt{\vartheta_{0}}\right) = 2.33$ which is greater than any payoff in Table.\ref{tab2}.
However, playing by optimal strategies against $1-\sqrt{\vartheta_{0}}$ is preferable for each agent $W_1\left(\eta'_1,1-\sqrt{\vartheta_{0}}\right)=2.73$ and $W_1\left(\eta''_1,1-\sqrt{\vartheta_{0}}\right)=2.77$, while 
$W_1\left(1-\sqrt{\vartheta_{0}},1-\sqrt{\vartheta_{0}}\right)=2.33$

The latter situation holds for any initial temperature ratio $\vartheta_{0}$.

Thus, both agents face the well-known prisoners dilemma game.

\subsubsection{Competition of two adaptive agents.}

Let us consider now the competition of locally adaptive heat engines. We will assume that $\eta_{k}=\alpha_{k}(1-\vartheta(t)),~k=1,2$, i.e the efficiencies are varying according to the environmental changes.  
Putting the efficiencies back into the (\ref{urd4}) we obtain 

\begin{eqnarray}
\label{jur}
\dot \vartheta=2\mu(1-\bar\alpha)\vartheta^{2}(1-\vartheta),
\end{eqnarray}

where $\bar \alpha=(\alpha_{1}+\alpha_{2})/2$. The only rest point of (\ref{jur}) is $\vartheta=1$. Thus, in contrast to the previous case, here both engines operate till the full consumption of the available  resources. The solution is found from (\ref{lamb2}) by changing $\mu \to 2\mu$ and $\alpha\to \bar\alpha$. 

The stored energy of each agent ($k=1,2$) is found by analogy of (\ref{ueda}) and is given by the following expression

\begin{eqnarray}
\label{ued6}
-\int_0^\infty {\rm d}t\, \hat J^{[k]}_{21}(t)
=\frac{\alpha_{k}(1-\alpha_{k})}{2 \mu (1-\bar\alpha)}\bigg(\frac{1}{\vartheta_0}-1+\ln{\vartheta_{0}}\bigg)
\end{eqnarray}

From (\ref{ued6}) it follows that those engines for which $\alpha=1/2$ i.e $\eta(t)=\frac{1-\vartheta(t)}{2}$ will outperform others. Indeed, according to (\ref{eq:1b}) an engine operating by the $\alpha=1/2$ level is maximizing its power $\hat J_{21}$ at each moment in time and eventually maximizes the stored energy due to the absence of any time constraints: since both engines are operating as long as $\vartheta(t) \neq 1$.

Let us discuss competition between two types of heat engines-- operating close to the maximum efficiency $\alpha_{me} \to 1$
and operating at maximum power $\alpha_{\rm mp}=\frac{1}{2}$,  for the given initial temperature ratio $\vartheta_{0}$. 

From (\ref{ued6}) follows that the stored energy of each of the agents operating by near maximum efficiency level $\alpha_{\rm me}\to1$ will be proportional to $\frac{\alpha_{\rm me}}{2\mu}$. While, the stored energy of the agents operating at maximum power regime is proportional to $\frac{\alpha_{\rm mp}}{2\mu}$, note that the prefactor is the same for both cases. Due to $\alpha_{\rm me}>\alpha_{\rm mp}$ it is better to operate at higher efficiency level if opponent does so. 

However, it is better to operate at the maximum power regime $\alpha_{\rm mp}$ when opponent is operating at $\alpha_{\rm me}$, since $\alpha_{\rm mp}(1-\alpha_{\rm mp})>\alpha_{\rm me}(1-\alpha_{\rm me})$ (the denominator in both expression is same $2\mu\left(1-\frac{\alpha_{\rm mp}+\alpha_{\rm me}}{2}\right)$).

Thus, adaptive agents are facing the same dilemma as non-adaptive agents.

\subsubsection{Adaptive v.s. non-adaptive agent.}

Here we discuss competition of two agents-- one with fixed internal structure (non-adaptive) and adaptive agents. In contrast to the above discussed competition scenario, here, only the agent with fixed internal structure has a finite lifetime. Indeed, the temperature ratio dynamics has the following form 

\begin{eqnarray}
\label{fa1}
\dot{\vartheta}=\mu\vartheta^2(1-\vartheta-\eta)+\mu\vartheta^2(1-\vartheta-\alpha(1-\vartheta))
\end{eqnarray}

where we have used (\ref{fox}) for the efficiency of the adaptive agent, and $\eta$ is an efficiency of the agent with  fixed internal structure.
The second term of (\ref{fa1}) nullifies only on the point $\vartheta=1$ , i.e. when the resources are fully depleted. While, the first term nullifies at the time $\vartheta(\tau)=1-\eta$. 

Solving (\ref{fa1}) by the analogy of the above mentioned procedure we found

\begin{eqnarray}
\label{fa2}
&&\vartheta(t)=a\left(1+W\left[ \left(\frac{a}{\vartheta_{0}}-1\right)e^{\frac{a}{\vartheta_{0}}-1-a\bar\mu t}\right]\right)^{-1},\\
&& a\equiv1-\frac{\eta}{2-\alpha}, \qquad \bar\mu\equiv(2-\alpha-\eta)\mu.\nonumber
\end{eqnarray}

The lifetime $\tau$ of the agent with fixed internal structure is found from (\ref{fa2}) and the condition $\vartheta(\tau)=1-\eta$.

\begin{eqnarray}
\label{fa3}
\tau=\frac{1}{a\bar\mu}\ln{\left[\left(\frac{a}{\vartheta_{0}}-1\right)e^{\frac{a}{\vartheta_{0}}-1}\frac{2-\alpha}{(1-\alpha)\eta}e^{-\frac{(1-\alpha)\eta}{2-\alpha}}\right]}
\end{eqnarray}

The stored energy of non-adaptive agent during its' lifetime will be equal to 

\begin{eqnarray}
\label{fa4}
\int_{0}^{\tau}\eta (1-\vartheta(t)-\eta) d t= \eta(1-\eta) \tau-\eta\int_{0}^{\tau}\vartheta(t) d t =  \eta(1-\eta) \tau-\frac{\eta}{\bar\mu}\ln{\left[\frac{\frac{a}{\vartheta_{0}}-1}{\frac{1-\alpha}{2-\alpha}\eta}\right]}
\end{eqnarray}

where the integral is found from (\ref{fa2}) and (\ref{fa3})

\begin{eqnarray}
\label{fa5}
&&\int_{0}^{\tau} \vartheta(t) d t= a \int_{0}^{\tau}\frac{d t}{1+W\left[ \left(\frac{a}{\vartheta_{0}}-1\right)e^{\frac{a}{\vartheta_{0}}-1-a\bar\mu t}\right]}=\nonumber\\
&&=\frac{1}{\bar\mu}\int_{\hat{t}(\tau)}^{\hat{t}(0)} \frac{d \hat{t}}{\hat{t} \left(1+W\left[ \left(\frac{a}{\vartheta_{0}}-1\right)e^{\frac{a}{\vartheta_{0}}-1}\hat{t}\right]\right)}=\frac{1}{\bar\mu}\ln{\left[\frac{\frac{a}{\vartheta_{0}}-1}{\frac{1-\alpha}{2-\alpha}\eta}\right]}
\end{eqnarray}

The stored energy of the adaptive agent is composed of two terms-- the one in which another agent has been in the competition and after the turning of that agent. 
Note that the power  of the adaptive agent has the following form $-\hat{J}_{21}=\alpha(1-\vartheta)(1-\vartheta-\alpha(1-\vartheta))$.
The stored energy of the adaptive agent during the competition  time $\tau$ is equal 

\begin{eqnarray}
\label{fa6}
&&\int_{0}^{\tau} \alpha(1-\alpha) (1-\vartheta(t))^2 d t= \frac{\alpha(1-\alpha)}{a\bar\mu}\int_{\hat{t}(\tau)}^{\hat{t}(0)} \frac{d \hat{t}}{\hat{t}} \frac{\left(1-a+W\left[\left(\frac{a}{\vartheta_{0}}-1\right)e^{\frac{a}{\vartheta_{0}}-1}\right]\right)^2}{\left(1+W\left[\left(\frac{a}{\vartheta_{0}}-1\right)e^{\frac{a}{\vartheta_{0}}-1}\right]\right)^2}=\\
\label{fa7}
&&=\frac{\alpha(1-\alpha)}{a\bar\mu}\left((1-2\alpha)\ln\left[\left(\frac{a}{\vartheta_{0}}-1\right)\frac{2-\alpha}{(1-\alpha)\eta}\right]+a^{2}\ln{\left[\frac{(1-\frac{\vartheta_{0}}{a})\zeta}{(1-\alpha)\eta}\right]}+\frac{a}{\vartheta_{0}}-\frac{\zeta}{2-\alpha}\right),
\end{eqnarray}

where we have denoted $\zeta\equiv 2-\alpha+(1-\alpha)\eta$. After switching of the agent with fixed internal structure the time evolution of temperature ratio is given by (\ref{lamb2}), where $\vartheta_{0}$ is substituted by $1-\eta$. The stored energy of the adaptive agent will be found from (\ref{ueda}) by substituting $\vartheta_{0}\to 1-\eta$

\begin{eqnarray}
\label{fa8}
\int_{\tau}^{\infty}\a(1-\a)(1-\vartheta(t))^2 d t= \frac{\a}{\mu}\left(\frac{1}{1-\eta}-1+\ln{1-\eta}\right)
\end{eqnarray}

\comment{

\section{Population dynamics model of Ref.~\cite{shuster}}

There are two populations with sizes $N-1$ and $N_2$, respectively. There is a food (resource) with size $S$:
\BEA
\label{ph1}
&&\dot{S}=v-N_1(t)J_1(S)-N_2(t)J_2(S),\\
\label{ph2}
&&\dot{N}_1=\eta_1N_1(t) J_1(S)-d N_1(t),\\
\label{ph3}
&&\dot{N}_2=\eta_2N_2(t) J_2(S)-d N_2(t),
\EEA
where $v$ is the food supply rate, $d$ is the decay rate in the populations, $J_1(S)$ and $J_2(S)$ are the resource intake rates. Here $\eta_1$ and $\eta_2$ are efficiency of the resource conversion into the population growth. For $S\to 0$ both $J_1(S)$ and $J_2(S)$ should go to zero sufficiently quickly. Then we shall certainly have $S(t)>0$ if $S(0)>0$. 

Eqs.~(\ref{ph1}, \ref{ph2}) imply:
\BEA
\frac{\d}{\d t}\,\frac{N_1}{N_2}=\frac{N_1}{N_2}(\eta_1J_1(S)-\eta_2J_2(S)).
\EEA
Thus the outcome of the competition between the two populations is determined solely by their power difference 
$\eta_1J_1(S)-\eta_2J_2(S)$. Now if $\eta_1J_1(S)>\eta_2J_2(S)$, then the second population is excluded in time, $N_2(t)\to 0$ for $t\to\infty$, while the size of the first population converges to a finite steady-limit $N_1(\infty)$. Now $N_1(\infty)$ is finite, since an infinite $N_1(\infty)$ would imply negative $S$, which is impossible, since $J_1(S)$ and $J_2(S)$ go to zero for $S\to 0$. 

Once $N_1(\infty)$ is finite, we find from (\ref{ph2}): $\eta_1J_1(S(\infty))=d$. Then (\ref{ph2}) implies: 
\BEA
N_1(\infty)=\frac{v\eta_1}{d}.
\EEA
Thus the competition is determined solely by the difference in powers, but once the population survived, its size is determined by the efficiency. In particular, the size of a single population (without competitors) is determined by the efficiency only and hence favors maximally large efficiencies. Note that the efficiencies in this model are not demanded to be smaller than one; nothing prevents arbitrary positive values for them. 

}
\end{document}